\documentclass[useAMS,usenatbib]{mn2e}

\usepackage{epsfig}

\def\al{$\rm[\alpha/Fe]$ }
\def\fe{$\rm[Fe/H]$ }

\def\ee#1{\times10^{#1}}

\title[Major Gas-Rich Disc-Disc Mergers]
{Structure, Kinematics, and
Chemical Enrichment Patterns after Major Gas-Rich Disc-Disc Mergers}
\author[S. Richard et al.]{Simon Richard,$^{1,2,3}$
Chris B. Brook,$^{4}$
Hugo Martel,$^{1,3}$
Daisuke Kawata,$^{5}$
\newauthor
Brad K. Gibson,$^{4}$ and
Patricia Sanchez-Blazquez$^{4,6,7}$\\
$^{1}$D\'epartement de physique, de g\'enie physique et 
d'optique, Universit\'e Laval, Qu\'ebec, QC, G1K 7P4, Canada\\
$^{2}$Hubert Reeves Fellow\\
$^{3}$Centre de Recherche en Astrophysique du Qu\'ebec\\
$^{4}$Jeremiah Horrocks Institute for Astrophysics \&
Supercomputing, University of Central Lancashire, Preston, PR1 2HE, UK\\
$^{5}$Mullard Space Science Laboratory, University College London,
Holmbury St. Mary, RH5 6NT, UK\\
$^{6}$Instituto de Astrof\'{\i}sica de Canarias, E-38200 La Laguna, Tenerife, Spain\\
$^{7}$Departamento de Astrof\'{\i}sica, Universidad de La Laguna, E-38205 La Laguna, Tenerife, Spain}
\begin{document}

\date{}

\pagerange{\pageref{firstpage}--\pageref{lastpage}} \pubyear{}

\maketitle

\label{firstpage}

\begin{abstract}
We used an N-body smoothed particle hydrodynamics algorithm, with a detailed treatment of star formation, supernovae feedback, and chemical enrichment, to perform eight simulations of mergers between gas-rich disc galaxies. We vary the mass ratio of the progenitors, their rotation axes, and their orbital parameters and  analyze the kinematic, structural, and chemical properties of the remnants. Six of these simulations result in the formation of a merger remnant with a disc morphology as a result of the large gas-fraction of the remnants. 
We show that stars formed during the merger (a sudden starburst occur in our simulation 
and last for  $0.2-0.3\,{\rm Gyr}$) and those formed after the merger have different kinematical and chemical properties. The first ones are  located in thick disc or the halo. They are partially supported by velocity dispersion and have high \al ratios even at metallicities as high as $\rm[Fe/H]=-0.5$. The former ones -- the young component -- are located in a thin disc rotationally supported and have lower \al ratios. The difference in the rotational support of both components results in the rotation of the thick disc lagging that of the thin disc by as much as a factor of two, as recently observed.
%We also find counter-rotating stars both in the old and young populations.
%A variety of structures are formed during the merger, i.e., most simulations form a ring of young stars
%and two simulations formed a bar.
%The scale-length of the thick disc is  either equal to the one of the thin disc, or larger by factors
%up to 1.60 and in six out of the eight simulations, 
%the thin and thick discs both have exponential luminosity profiles and are nearly coplanar.
We find  that, while  the kinematic and structural properties of the merger remnant depends
strongly upon the orbital parameters of the mergers, there is  a remarkable
uniformity in the chemical properties of the mergers. This suggests that
general conclusions about the chemical signature of gas-rich
mergers can be drawn.
\end{abstract}


\section{INTRODUCTION}

\subsection{Disc-Disc Mergers}

In a $\Lambda$ Cold Dark Matter Universe ($\Lambda$CDM), the formation 
of structures generally involves the merging of
smaller structures. Mass is built up predominantly by
the merger of objects of masses between $3-30\%$ of the galaxy's
total mass at redshift $z=0$ \citep{stewartetal08}. Since $z=2$,
approximately 60\% of galaxies have experienced a major 
merger, which we will take in this paper as
those involving a ratio of masses 3:1 or higher. 
Also, an important fraction of the baryonic mass (30 - 50\%)
of $z=0$ galaxies is provided by major mergers \citep{stewartetal09}. 
Recent observations show that nearly 20\% of the massive disc galaxies
 have undergone a major merger since $z\sim1.5$ \citep{lopezetal09}
This process is virtually scale-free, meaning that low-mass dark
matter halos, in which disc galaxies presumably reside,
have undergone similar merging histories to high-mass
dark matter halos in which, predominantly, ellipticals
reside. This at first is puzzling, as mergers have long
been associated with forming spheroidal structures, i.e.
early-type galaxies \citep{toomre77}. Following this line of
reasoning, every galaxy in the nearby universe should
be an elliptical or an irregular, which is clearly not the
case. Furthermore, observed elliptical galaxies can be
described by simple relations relating colors, luminosities, masses, etc. 
These simple relations are difficult to
reproduce with a scenario involving collisions \citep{bs98,bs99}.

The resolution of this, of course, lies
with the baryonic physics. 
Studies using two stable discs as initial conditions
(\citealt{sh05,robertsonetal06}; \citealt[][hereafter B07]{brooketal07};
\citealt{rb08})
as well as fully cosmological simulations \citep{gv08} have shown that binary
gas-rich mergers can result in disc morphologies.
Hence, major mergers between gas-rich disc galaxies can possibly play 
an important role in the formation of a large number of the
spiral galaxies we observe today. These facts bring an addition to the
original scenario for the
formation of disc galaxies, which involved the assembly
of a halo of dark matter and ionized gas by hierarchical clustering 
and mergers of sub-halos, followed by the
collapse and virialization of the dark matter halo, and finally 
the dissipative collapse of the baryonic component,
resulting in a rotationally-supported disc, that will later
fragment to form stars \citep{wr78,fe80,blumenthaletal84}. 

One must also
consider a formation scenario involving major collisions
between spiral galaxies. The first numerical simulations of major 
collisions between disc galaxies were performed under the assumption
that dissipation by gasdynamical effects was negligible.
The results showed that the remaining galaxy had several properties 
attributed to elliptical galaxies, including luminosity profiles 
\citep{nb03,barnes02,bh92,hernquist92,hernquist93,bs98,bs99,nbh99}.
Even with the quite recent inclusion of feedback processes, the 
results remained mostly
unchanged \citep{bh96,mh96,springel00,naabetal06}. Minor collisions have also
been studied in the past and results showed that this kind
of collision tends to form galaxies whose disc is destroyed
or very perturbed \citep{hm95,bekki98}.
All the aforementioned studies assumed 
a ratio gas/stars more representative of present low redshift mergers.
 Of course, stars have not always been present inside
galaxies. Going back in time, we can expect to find
fewer and fewer stars, and therefore a more important
gaseous component \citep{stewartetal09}.  

\subsection{Thick Discs} 

The Milky Way's thick disc is made predominantly of old
stars, with ages in the range of $8-12\,{\rm Gyr}$, 
significantly older than the typical star found in the thin disc
\citep{gwj95}. The thick and thin disc differ in their structural,
chemical, and kinematical properties: (1) The thick disc
has a scale height of about 900 pc, compared to 300 pc for
the thin disc \citep{juric08}. (2) The stars in the thick disc have a low
metallicity, $\rm[Fe/H]\sim-0.6$ on average, and a ratio \al
larger than the stars in the thin disc. (3) Stars in the
thick disc tend to rotate slower than stars in the thin
disc, by about $40\,{\rm km\,s^{-1}}$ \citep{gwk89} and have an
average vertical velocity dispersion of $45\,{\rm km\,s^{-1}}$ compared
to $\sim10-20\,{\rm km\,s^{-1}}$ for the stars in the thin disc 
\citep{delhaye65,cb00}. 

The formation of the Milky
Way's thick disc remains an open question, with popular
theories including the heating of a thin disc by minor
mergers \citep{quinnetal93,kazantzidisetal08,kzb08}, or
the direct accretion of stars, which are tidally torqued
into the plane of the existing thin disc \citep{abadietal03},
or dynamical heating by destruction of very massive star
clusters \citep{kroupa02,ee06}.
We favor a scenario in which the Milky Way's
thick disc is formed during the high-redshift epoch in
which mergers are most common in $\Lambda$CDM, and accretion rates are high, 
with the disc which formed during
this turbulent period being born thick; the thin disc subsequently 
grows during the relatively quiescent period
which follows, which in the case of the Milky Way is approximately 
the last 10 Gyrs \citep{brooketal04b,brooketal05,brooketal06}.
%Another interesting suggestion
%is that the stars in the young thin disc are dynamically
%heated by the destruction of very massive star clusters \citep{kroupa02,ee06}.

Recent observations suggest that all spiral galaxies are
surrounded by a red, flattened envelope of stars, and
these structures have been interpreted as showing that
thick discs are ubiquitous in spiral galaxies \citep{db02}. 
Hence, understanding the formation
of the thick disc will provide major insights into the formation 
processes of spiral galaxies. 
If we make the hypothesis that thick disc are formed during gas-rich
major mergers, it would mean that a significant number of spiral galaxies
have experienced major mergers.
%As significant numbers of spiral galaxies 
%have undergone major mergers, the
%disc galaxies which form in gas-rich mergers must necessarily have an 
%associated thick disc. 
%Previous studies of collisions and mergers between gas-rich galaxies
%\citep{robertsonetal06,coxetal08,jnb09} included the effect of star formation
%and supernova feedback,
Indeed, simulations
have shown that gas-rich mergers tend to form two disc 
components (\citealt{robertsonetal06}; B07). These simulations
included the effect of star formation and supernovae feedback, 
which allowed to study the kinematical and structural properties of 
galaxy merger remnants in detail. However, they 
did not include a detailed treatment of chemical enrichment and,
therefore,  could not predict the abundances of metals in stars and the
interstellar medium \citep[but see also][who studied the formation of 
ellipticals galaxies due to a more violent major merger]{bs98,bs99} . 
It is well-established
that the thick disc of the Milky Way possesses a unique chemical
signature, which is believed to be directly related to the formation
process and it is important to analyze if the merger scenario for the 
formation of the thick disc component is compatible with  these chemical 
signatures. This is the basis of our study here: we believe that
the remnants of major mergers may be associated with
a significant fraction of the ``red envelopes'' observed in
\citet{db02}, i.e. with extragalactic thick discs. Since we expect the 
conditions for merging galaxies during hierarchical clustering at a 
high redshift to vary, we need to
consider a wide range of parameters, such as mass ratio,
orbit, and gas fraction of the merger progenitors. In this
paper, we present a series of 8 simulations of mergers
of gas-rich disc galaxies, with different mass ratios and
orbital parameters. We analyze the kinematic and structural properties 
of the merger remnants. The results for
one of the simulations have already been presented in a
previous paper (B07). We now complete this
work by presenting the entire suite of simulations.

For kinematic properties, we determine the relative 
importance of rotation and velocity dispersion in 
providing support. We also investigate if
a counter-rotating component arises naturally in a gas-rich merger, as 
such components are observed in nature \citep{yd06}.
For structural properties,
we determine if the luminosity profiles of these thick discs
formed in major mergers follow an exponential law, and
if so, what their respective scale lengths are. We determine if
the thick and thin discs in merger remnants are coplanar, or
if there is a significant angle between them, something that
was not included in our previous work.

However, the main goal of the present paper is to study the viability 
of the merger mechanism for the formation of the thick disc from the 
chemical evolution's point of view, but without neglecting the 
kinematical and dynamical properties. 
%Disc formation after a gas-rich merger is still a matter of%
%debate, but our simulations with a completely independent code 
%provide support for the results of \citet{robertsonetal06} and others.
With this in mind, we
choose initial conditions likely to produce a discy remnant.

In this paper, we use the numerical algorithm GCD+, which includes a
detailed treatment of chemical enrichment, to investigate whether
key abundance ratios such as \al and \fe in simulated remnants of 
gas-rich disc galaxy
mergers have values and distributions similar to the ones found
in the observed disc galaxies, including the Milky Way.

The reminder of this paper is organized as follows: in
\S2, we briefly describe the numerical simulations employed, including
the N-body/SPH code adopted (GCD+), the initial conditions software
(GalactICS), and the basic parameters of our 8 realizations.
The kinematical and structural properties of the merger remnants are 
presented in \S3 and discussed in \S4. Conclusions are presented in \S5.

\section{THE NUMERICAL SIMULATIONS}

\subsection{The Algorithm}

All simulations were performed using GCD+ \citep{kg03a,kg03b}, 
self-consistently modeling
the effects of gravity, gas dynamics, radiative cooling, and star
formation. GCD+ is a Tree/SPH algorithm that
includes Type~Ia and Type~II SNe feedback, and traces 
the lifetimes of individual stars, enabling us to  monitor the
chemical enrichment history of our simulated galaxies. Star formation occurs
in a convergent gas velocity field where gas density is greater than a
critical density, $\rho_{\rm crit}= 2 \times 10^{-25}{\rm g\,cm^{-3}}$. 
The star formation rate (SFR) of eligible gas particles is then
$d\rho_*/dt=d\rho_g/dt=c_*\rho_g/t_g$
where $c_*=0.05$ is a dimensionless star formation efficiency,
and $t_g$ is the dynamical time. This formula corresponds to a
Schmidt law: SFR $\propto\rho^{1.5}$. The mass, energy, and heavy
elements are smoothed over the neighboring gas particles using the SPH
smoothing kernel. The code separately tracks
the abundances of 9 elements: H, He, C, N, O, Ne, Mg, Si, and Fe.
Gas within the SPH smoothing kernel of Type~II SNe explosions is prevented
from cooling, creating an adiabatic phase for gas heated by such SNe. This
adiabatic phase is assumed to last for the lifetime of the lowest mass star
that ends as a Type~II SNe, i.e., the lifetime of an $8M_\odot$ 
star (100 Myr). This is similar to a model presented in \cite{tc00}. 
Stellar yields are dependent of the progenitor mass and metallicity. 
For Type~II SNe, we
use the metallicity-dependent stellar yields of \cite{ww95}
for stars with mass over $11M_\odot$, while for
low- and intermediate-mass stars, we use the stellar yields of
\cite{vdhg97}. 
For Type~Ia SNe, we adopt the model of \cite{ktn00} in which Type~Ia SNe 
occur in binary systems with a primary star and a companion of a definite 
mass and metallicity range. The yields are from \cite{iwamotoetal99}.
Note that GCD+ relaxes the instantaneous recycling approximation
and takes into account the lifetime of progenitor stars, and
chemical enrichment from intermediate mass stars.
The code calculates, within each timestep, 
the amount of energy and heavy element released by every star particle 
and distributes them over the neighbours using the SPH kernel
for both Type II and Ia SNe.
We did not include in the present paper any black hole feedback.
Several studies have demonstrated that this can be very important
for the final morphology and the  termination of the star formation in 
the merger remnant \citep{sdh05,oka05,jnb09}. However, the implementation 
of the blackhole feedback in numerical simulations
 is still highly ambiguous and we prefer here, for simplicity, 
to include only SN feedback.
For more details about GCD+, we refer the reader to \citet{kg03a,kg03b,
brooketal04a}.

\subsection{Setting Up Initial Conditions}

\begin{figure*}
\includegraphics[width=6in]{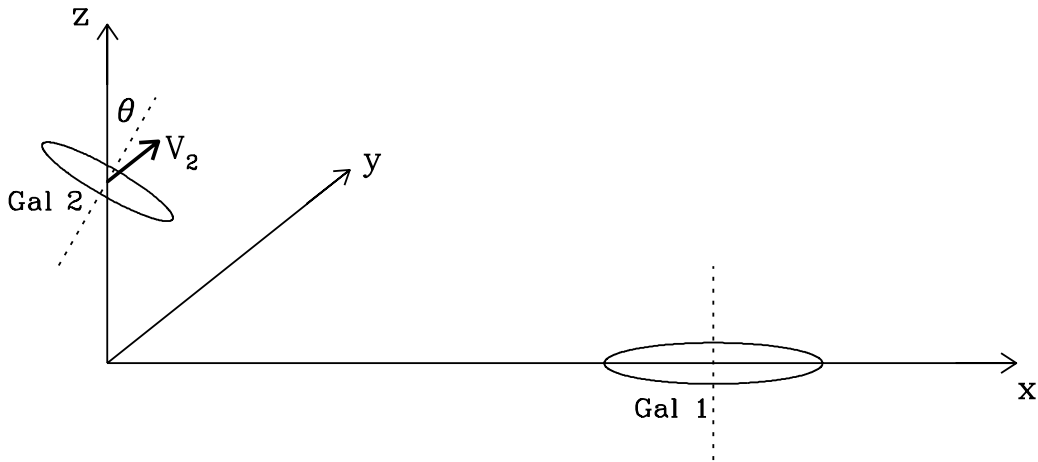}
\caption{Geometry of the initial conditions. The $Y$-axis is pointing away 
from the viewer. Gal1 is initially at rest, while Gal2 is initially
moving away from the viewer, in the $+Y$ direction, as the thick arrow
shows. The dotted
lines indicate the rotation axes of the galaxies. 
Gal1 is in the $X-Y$ plane, with
the rotation axis in the $Z$-direction. The rotation axis of Gal2 is
in the $X-Z$ plane, at an angle $\theta$ relative to the $Z$-axis.
Both galaxies are rotating clockwise when seen from above, hence
the left edges are moving away from the viewer, and the right edges are moving
toward the viewer.
The particular case shown on this figure corresponds to simulation M12,
with the edges of the discs located at two scale-lengths.
\label{fig-initial}}
\end{figure*}

In each simulation, the initial conditions consist of
two galaxies with exponential gas discs embedded in dark matter halos.
These galaxies are generated using the GalactICS package
\citep{kd95},
producing stable discs for a large number of galactic rotations.
For the dark matter halos, GalactICS use the lowered Evans 
model \citep{evans93}, 
which leads to a constant-density core. This model differs 
from models which have a steeper central density profile
suggested by $\Lambda$CDM cosmological simulations
\citep{nfw96,nfw97,mooreetal99,ghignaetal00,js00,klypinetal01}. 
For our purpose this difference is not critical, as the central regions
contain only a small fraction of the total mass. The star formation mechanism
 is shut down for the first $\sim180{\rm Myr}$ enabling very gas-rich mergers.  
In our simulations, the gas ratio immediately before merger is more 
relevant than the initial value.
Pure gas disc will induce an unrealistic starburst initially. However,
by turning off star formation initially,
we can control the gas fraction at the merger epoch.
For each simulation, the most massive galaxy (Gal1) has total mass 
of $5\times10^{11}M_\odot$ (except for simulation M11) and a scale 
length of 4.5~kpc. The mass and scale length of the second galaxy (Gal2)
depends on the specific mass ratio chosen for the simulation.  
Using the observed gas fraction of galaxies at high \citep{erbetal06} and low
\citep{mcgaugh05} redshift, 
\citet{stewartetal09} provide a relation between gas
fraction and redshift for galaxies of various masses. Using this relation
(their equation~[1]) allows us to estimate that the mergers in the
epoch between $z=2.37-3.7$ would be between galaxies which have gas
fractions as high as the ones used in our study. Note that this is
an upper bound of such epoch due to the fact that the high redshift
observations of \citet{erbetal06} are considered lower bounds of the gas
content. Also, these redshifts coincide with the epoch of formation of 
the stars in the Milky Way thick disc.

 %By combining equation~(1) of \cite{stewartetal09} with observational data 
%from \cite{erbetal06} we can estimate the redshift of 
%our progenitors. The results
%depend mostly on the stellar mass and the gas/star mass ratio of the 
%progenitor. This is an upper limit ,
%considering that the Schmidt law may underestimate the total gas content. 
%For our sample, this upper limit fall in the range $z=2.37-3.7$.

The initial configuration of the system is illustrated in 
Fig.~\ref{fig-initial}. 
Gal1 is initially located 
at ${\bf R}_1=(50,0,0)\,{\rm kpc}$, with no initial velocity,
and its rotation axis pointing in the $Z$ direction. Gal2
is located at ${\bf R}_2=(0,0,15){\rm kpc}$ (except for simulations M12C and 
M12orb), with
initial velocity ${\bf V}_2=(0,100,0){\rm km\,s^{-1}}$ (except for 
simulation M12z),
and its rotation axis in the $X-Z$ plane at an angle $\theta$
relative to the $Z$-axis. We shall refer to this angle as the
{\it interaction angle}.
All encounters are prograde-prograde, except for simulation rM12, for
which the rotation of Gal2 is retrograde.

\subsection{A Series of 8 Gas-Rich Mergers}

\begin{table*}
 \centering
 \begin{minipage}{220mm}
  \caption{Initials Conditions for All Simulations\label{table-initial}}
  \begin{tabular}{@{}lccccccccc@{}}
  \hline
    Run & $M_{\rm Gal1} (M_\odot)$ &
    mass ratio & $f_{\rm gas}$ & $\theta$ &
    $V_{2z}\;{\rm(km\,s^{-1}})$ & ${\bf R}_2\;{\rm(kpc)}$ 
    & $e$ & $\omega$[degree] & $\iota$[degree] \\
  \hline
 M12     & $5.0\ee{11}$ & 2:1  & 0.94 &  30.0 &   0 & 0,  0, 15 & 0.79 & 90 & $-13$ \\
 M12orb  & $5.0\ee{11}$ & 2:1  & 0.67 &  30.0 &   0 & 0, 50, 15 & 0.86 & 51 & $-13$ \\
 M12z    & $5.0\ee{11}$ & 2:1  & 0.83 &  30.0 & 100 & 0,  0, 15 & 0.66 & 29 & $-20$ \\
 M1290   & $5.0\ee{11}$ & 2:1  & 0.96 &  90.0 &   0 & 0,  0, 15 & 0.91 & 90 & $73$\\
 rM12    & $5.0\ee{11}$ & 2:1  & 0.91 & 210.0 &   0 & 0,  0, 15 & 0.79 & 90 & $167$\\
 M11     & $2.5\ee{11}$ & 1:1  & 0.88 &  30.0 &   0 & 0,  0, 15 & 0.89 & 90 & $-13$\\
 M13     & $5.0\ee{11}$ & 3:1  & 0.92 &  30.0 &   0 & 0,  0, 15 & 0.70 & 90 & $-13$\\
 M110    & $5.0\ee{11}$ & 10:1 & 0.85 &  30.0 &   0 & 0,  0, 15 & 0.37 & 90 & $-14$\\
   \hline
  \end{tabular}
 \end{minipage}
\end{table*}

We have performed a total of 8 simulations with different initial
conditions, in which each galaxy is modeled using
100,000 dark matter halo particles and 40,000 gas particles, for a
total of 280,000 particles. This resolution is lower by a factor 2 than 
the one used in \cite{robertsonetal06}. Since the primary goal of this work 
is to experiment on chemical enrichment following a major merger, we had to
make some choices relatives to the numerical resolution we had to use. 
The primary concern being the relatively high computation time needed to 
distribute enrich gas to the particles in the neighborhood.
The baryon/DM mass fraction is 17\%, which is equal to the universal ratio
$\Omega_b/\Omega_0$ according to recent estimates 
\citep{wmap3,komatsuetal09}.
Every simulation starts with an initial
metallicity $\log(Z/Z_{\odot})=-4$ and $\alpha$-element
abundance $\rm[\alpha/Fe]=0.30$.

Table~\ref{table-initial} lists
the parameters used for generating the initial conditions for
each simulation.
The five simulations within the M12 ``family''
involve a major merger of two galaxies with a mass ratio of 2:1.
We start with M12, which has already been discussed in detail in B07. 
The galaxies collide after 320~Myr, and by that time 6\% of the gas 
has already been converted to
stars, leaving a gas fraction $f_{\rm gas}=0.94$.
The next 4 simulations within that family are variations 
of the ``parent'' simulation M12. 
In simulation M12orb, we increase the
initial separation between the galaxies, and in 
simulation M12z, we add an additional
velocity component in the $z$-direction. 
In both cases, the merger is delayed, and this
results in lower gas fractions at the time of the merger. 
In simulation M1290 we 
changed the interaction angle to 90 degrees. 
Simulation rM12 is similar to simulation M12,
except that the direction of rotation of Gal2
is reversed, making it retrograde relative to the orbit.

With the last 3 simulations, we consider different mass ratios. 
In simulation M11,
we used equal-mass galaxies. Also, this simulation is the only
one for which we use a smaller mass for
Gal1: $2.5\times10^{11}M_\odot$ instead of $5\times10^{11}M_\odot$.
In simulation M13, we use a mass ratio of 3:1. This is still
considered to be a major merger. In the simulation M110,
we use a mass ratio of 10:1, which corresponds to a minor merger.

Column 8 of Table~1 gives the initial eccentricities
of the orbits. Except for simulation M110
(the minor merger), the eccentricities are quite high,
corresponding to elongated elliptical orbits. As a result, the pericenters
are smaller than the actual size of the galaxies, and the collision takes
place before a full orbit is completed. 
Column~9 gives the angle between the position of the pericenter
and the line of nodes (intersection between the orbital plane and the plane 
of Gal1; see p.~632 of \citealt{tt72}).
Column~10 gives the inclination $\iota$ of Gal2 relative to the plane of the 
orbit. 

For the simulation M11, the softening length of the algorithm 
is $0.65\,\rm kpc$ for the dark matter and $0.48\,\rm kpc$ for the
baryons. For the other simulations, the softening length
is $0.82\,\rm kpc$ for the dark matter and $0.61\,\rm kpc$ for the
baryons. 

\begin{table}
 \centering
 \caption{Lookback Times and Final Gas Fraction\label{table-times}}
 \begin{tabular}{@{}lccc@{}}
  \hline
Run & $t_f$ [Gyrs] & $t_{\rm coll}$ [Gyrs] & $f_{\rm gas}^{\rm final}$\\
  \hline
 M12    & 2.50  & 0.27 & 0.14 \\
 M12orb & 3.40  & 0.83 & 0.18 \\
 M12z   & 2.50  & 0.50 & 0.20 \\
 M1290  & 2.50  & 0.25 & 0.10 \\
 rM12   & 2.47  & 0.32 & 0.18 \\
 M11    & 2.50  & 0.40 & 0.06 \\
 M13    & 2.50  & 0.32 & 0.18 \\
 M110   & 2.50  & 0.42 & 0.30 \\
  \hline
 \end{tabular}
\end{table}

Table~\ref{table-times} lists the time $t_f$ elapsed between the beginning
and the end of the simulation, the time $t_{\rm coll}$ elapsed between the
beginning of the simulation and the collision, and the final gas
fraction $f_{\rm gas}^{\rm final}$. In each simulation, the collision leaves
a very complex and chaotic system, which then relaxes to a ``quiescent state''
where the final structure and kinematics are well-established,
and the star formation rate has dropped significantly. We stopped the 
simulations after $2.5\,\rm Gyrs$, when the quiescent state had been reached.
One simulation (M12orb) was extended to 
$t_f=3.40\,\rm Gyrs$, which enabled us to check that the final state 
was indeed quiescent. Comparing the values of $f_{\rm gas}^{\rm final}$
with the values of $f_{\rm gas}$ listed in Table~\ref{table-initial} reveals
the efficiency of the starburst in converting gas to stars. The gas 
fraction typically dropped from $\sim90\%$ to $\sim15\%$. 
It is interesting to note that simulation
M110, the only case of a minor merger, was less efficient
in converting gas to stars than the other simulations, with
$f_{\rm gas}^{\rm final}=0.297$.

\section{RESULTS}

 Following the approach of \citet{robertsonetal06},
we have computed the SFR for all the simulations,
and used these results to identify stellar populations.
Some examples of them are shown in the top left panels of
Figs.~\ref{M12_kin}--\ref{M13_kin}.
Notice that the simulations start at $t=0$. 
The results are quite similar for all simulations.
A major starburst invariably occurs during the merger.
This starburst peaks between $\sim60M_\odot\,{\rm yr}^{-1}$ 
and $500M_\odot\,{\rm yr}^{-1}$, comparable to that seen in high-redshift
Lyman-break galaxies \citep{erbetal06}. 
The results are qualitatively similar to recent work 
involving similar gas ratio
 but different feedback assumptions  \citep{jnb09} or different gas ratio and
different feedback \citep{coxetal08}. The main quantitative difference 
come from the amount of gas available during the merger. As stated
in \citet[][see their Figs.~1 and 3]{jnb09}, there is a huge difference
in the maximum value of the SFR depending on gas mass fraction. 
The SFR is also dependent of the numerical methods used for feedback in case 
of major mergers \citep{coxetal06}.
After the starburst, the star
formation steadily drops, as the system relaxes and a disc forms.
We identify the beginning and the
end of the merger with the beginning and the end of the starburst,
respectively. We then define two populations of stars: {\it old stars},
which include stars already present in the galaxies before the merger
and stars formed during the merger by the starburst, and {\it young stars},
which include all stars formed after the merger, when the starburst is
completed. The definitions for young and old stars remain the 
same throughout the paper. 
The dashed vertical lines in the top-left panels
of Figs.~\ref{M12_kin}--\ref{M13_kin} indicate
the beginning and end of the starburst, identified by eye
as the merger boundaries are not crucial for the subsequent analysis. 
The time when the starburst begins
is listed in Table~\ref{table-times}. For most simulations, the starburst
lasts 0.2--0.3~Gyr. The simulation M12z (Fig.~\ref{M12z_kin})
is a notable exception,
with a starburst that is weaker and lasts for 0.5~Gyr.

We use the term {\it remnant} to designate the final state of each
simulation. 
Our simulations are meant to represent the interactions occurring at high
redshift, when the gas content is very high. In this sense, the
merger remnant represents objects at $z\sim2-3$ and the galaxies still have
time to form more stars until $z=0$.
The subsequent infall of extragalactic gas 
may also produce more stars at $z=0$ and help to reform a 
disc \citep{brooksetal09}.

\subsection{Structure and Density Profiles}

\begin{figure*}
\begin{center}
\includegraphics[width=5.5in]{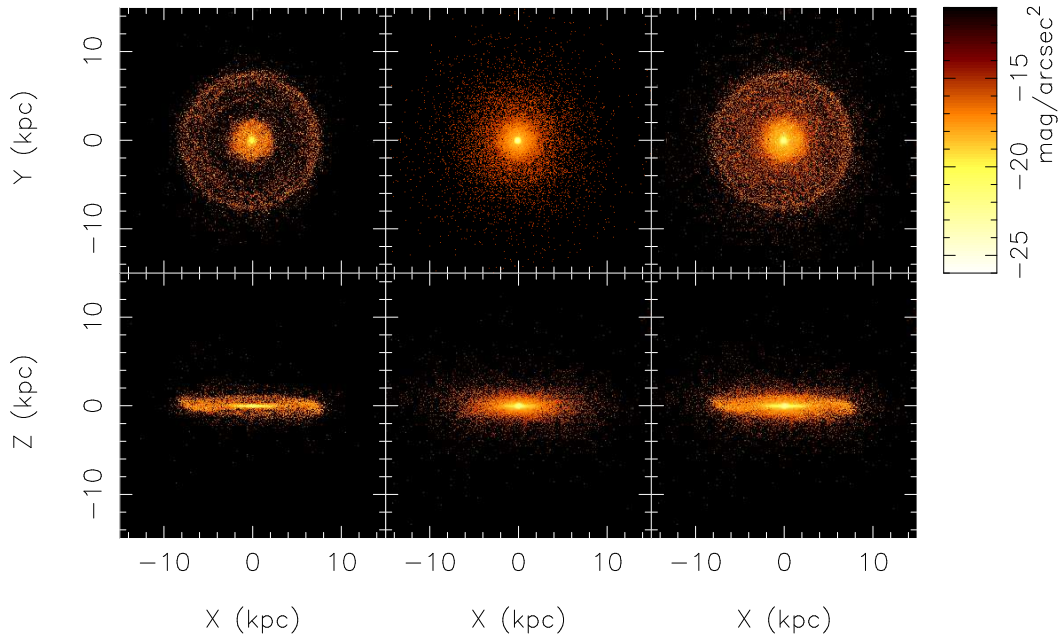}
\caption{V-band image of M12 seen face-on (top panels) and
edge-on (bottom panels), for young stars (left panels), old
stars (middle panels), and all stars (right panels). $X$, $Y$, $Z$
represent cartesian coordinates, with $Z$ along the axis of rotation.
}
\label{M12}
\end{center}
\end{figure*}

%\begin{figure*}
%\begin{center}
%%\includegraphics[width=5.5in]{M12orbB.ps}
%\includegraphics[width=5.5in]{figure03.eps}
%\caption{V-band image of M12orb seen face-on (top panels) and
%edge-on (bottom panels), for young stars (left panels), old
%stars (middle panels), and all stars (right panels).
%}
%\label{M12orb}
%\end{center}
%\end{figure*}

\begin{figure*}
\begin{center}
\includegraphics[width=5.5in]{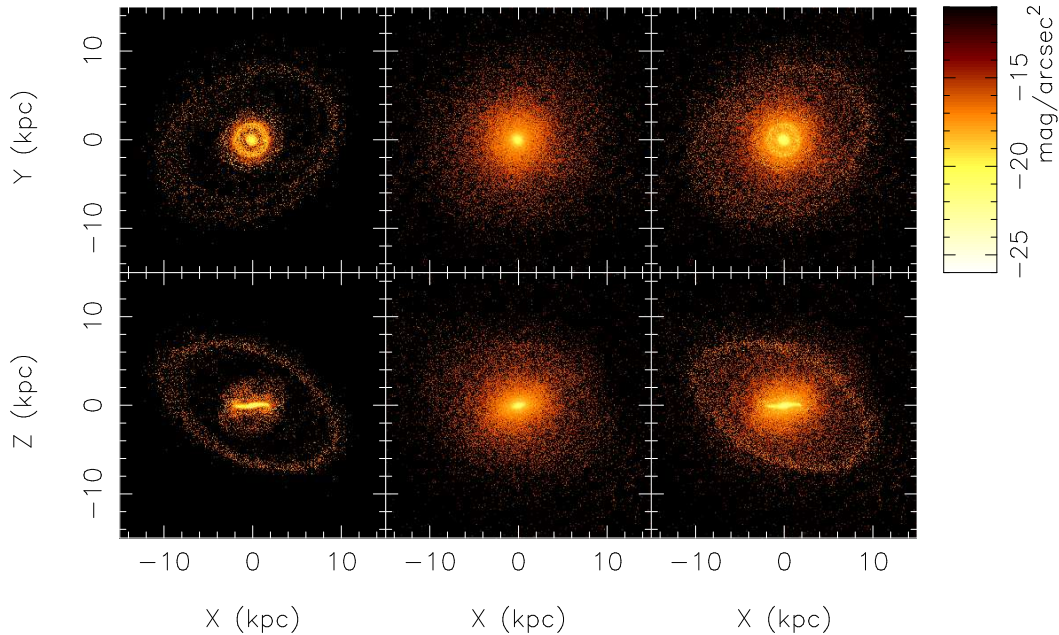}
\caption{V-band image of M12z seen face-on (top panels) and
edge-on (bottom panels), for young stars (left panels), old
stars (middle panels), and all stars (right panels). The gas shows a complex
structure that lasts for several hundred million years.
}
\label{M12z}
\end{center}
\end{figure*}

\begin{figure*}
\begin{center}
\includegraphics[width=5.5in]{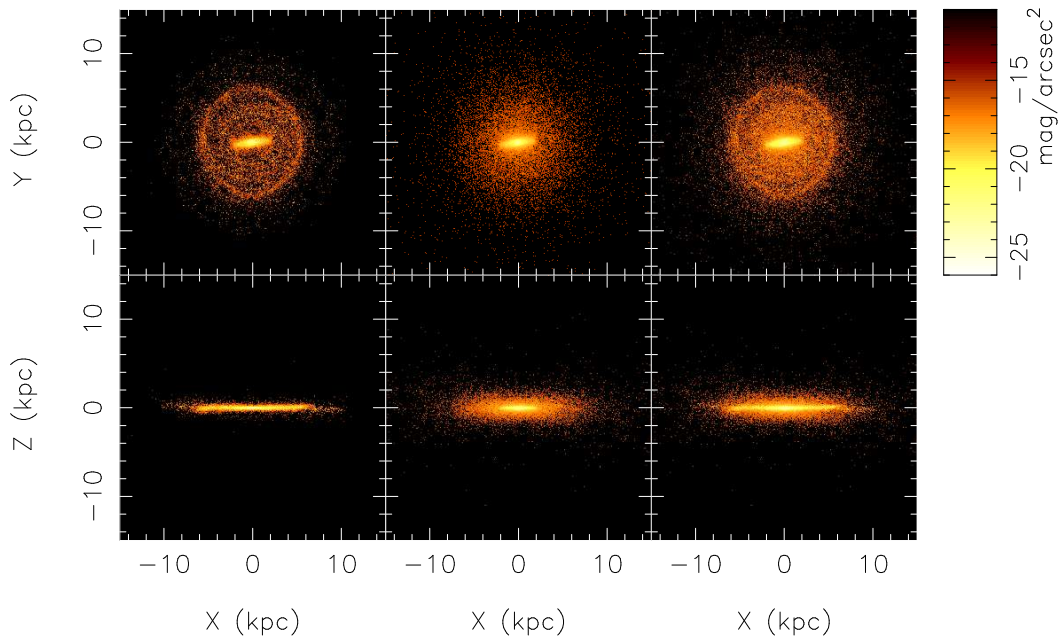}
\caption{V-band image of M1290 seen face-on (top panels) and
edge-on (bottom panels), for young stars (left panels), old
stars (middle panels), and all stars (right panels). For this simulation,
the stellar ring and the disc are not coplanar.
}
\label{M1290}
\end{center}
\end{figure*}

%\begin{figure*}
%\begin{center}
%%\includegraphics[width=5.5in]{rM12B.ps}
%\includegraphics[width=5.5in]{figure06.eps}
%\caption{V-band image of rM12 seen face-on (top panels) and
%edge-on (bottom panels), for young stars (left panels), old
%stars (middle panels), and all stars (right panels).
%}
%\label{rM12}
%\end{center}
%\end{figure*}

%\begin{figure*}
%\begin{center}
%%\includegraphics[width=5.5in]{M11B.ps}
%\includegraphics[width=5.5in]{figure07.eps}
%\caption{V-band image of M11 seen face-on (top panels) and
%edge-on (bottom panels), for young stars (left panels), old
%stars (middle panels), and all stars (right panels).
%}
%\label{M11}
%\end{center}
%\end{figure*}

%\begin{figure*}
%\begin{center}
%%\includegraphics[width=5.5in]{M13B.ps}
%\includegraphics[width=5.5in]{figure08.eps}
%\caption{V-band image of M13 seen face-on (top panels) and
%edge-on (bottom panels), for young stars (left panels), old
%stars (middle panels), and all stars (right panels).
%}
%\label{M13}
%\end{center}
%\end{figure*}

%\begin{figure*}
%\begin{center}
%%\includegraphics[width=5.5in]{M110B.ps}
%\includegraphics[width=5.5in]{figure09.eps}
%\caption{V-band image of M110 seen face-on (top panels) and
%edge-on (bottom panels), for young stars (left panels), old
%stars (middle panels), and all stars (right panels).
%}
%\label{M110}
%\end{center}
%\end{figure*}

\begin{figure*}
\begin{center}
\includegraphics[width=5.5in]{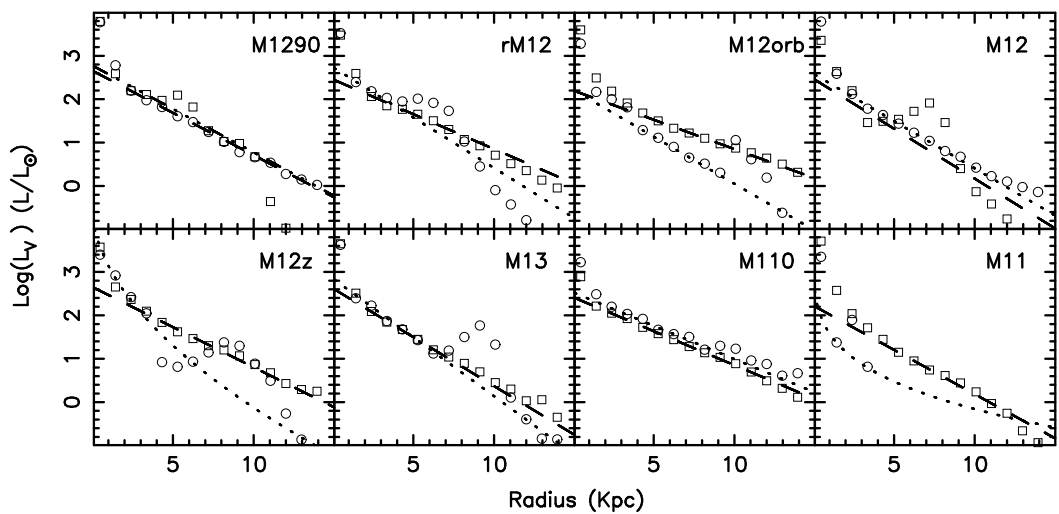}
\caption{Radial luminosity profile in the V-band, for old stars
(squares) and young stars (circles).
The straight lines show exponential fits of the old stars
(dashed) and young stars (dotted). The dotted
curves for simulation M12z and M11 are fits to a de~Vaucouleurs
profile for the young stars.
}
\label{allprof}
\end{center}
\end{figure*}

 At the end of each simulation, we calculated the V-band
luminosity of the merger remnant, using
the stellar synthesis model of \citet{ka97}.
From this, we produced mock V-band luminosity maps
(modulo the lack of dust extinction). 
Figs.~\ref{M12}--\ref{M1290} show the luminosity maps for
3 remnants. On each figure, the 
left panels show the stars born after the merger (young stars), 
the middle panels show the stars born before or during the merger (old stars)
and the right panels show all stars.

 We immediately see that old and young 
stars have very different distributions. In simulation M12
(Fig.~\ref{M12}), young and old
stars form discs that have comparable radii, but the edge-on views
(bottom panels) clearly shows that the young disc is thin, while the
old disc is thick. The presence of two distinct discs, a thin one and a 
thick one, is in remarkable agreement with observations \citep{yd06}, and this
motivated us to publish the results of that particular simulation in an earlier
paper (B07). We found that all the simulations except for simulations
M12z and M11 do result in the formation of a thin
disc made of young stars, and a thick disc made of old stars that were
already present before the merger, or formed during the mergers. These
stars end up either in the thick disc or in the halo. 

Simulations M12z and M11 show more elliptical like remnants.
Simulation M12z (Fig.~\ref{M12z}) 
produced a remnant that has a very 
complex structure. Even though we find a small disc made of young stars,
the overall structure resembles more an elliptical galaxy
than a disc galaxy. 
Simulation M11 also shows a small young disc.
Compared with the other simulations, the initial galaxies
Gal1 and Gal2 are less massive in simulation M11, 
but the intensity of the starburst taking
place during the merger is comparable. As a result, there is less gas
available after the merger to form young stars and build up an extended
thin disc. 

In most simulations, the merger resulted in the formation of a ring 
made of young stars. This implies that prograde gas-rich major mergers have
a tendency to form rings. These rings seem to be signatures of a gas 
substructure since most of the rings are coplanar with the thin disc. 
The large impact parameter use here is quite contrary to the 
normal scenarios forming a ring remnant  \citep{hernquistandweil93,mapelli08}. 
That can be a signature of a gas-rich merger,
and more work is needed to certify this assertion.
The simulation M12z formed a very complex structure, with a non-coplanar
ring connected to a very small thin disc by spiral arms, embedded in
an ellipsoid distribution of old stars (Fig.~\ref{M12z}).
Also, the face-on views of the mergers for simulations M1290 and
rM12 (top panels of Figs.~\ref{M1290} for an example
% and \ref{rM12}
) clearly reveal
the presence of a central bar made of both old and young stars. No bars
are seen in the other simulations.

We can use the surface density brightness to trace the radial 
luminosity profiles of the discs. 
The radial luminosity profiles
are shown in Fig.~\ref{allprof}, for all simulations. We calculated
these profiles separately for young stars and old stars.
The dotted and dashed
lines are fits to the profiles of young and old stars, respectively.
In cases where a ring was clearly visible, we excluded it from the fit.
All profiles are well-fitted by an exponential profile, except for
simulations M12z and M11, where the young stars are better-fitted by a
de~Vaucouleurs profile. We can use the exponential profiles to
calculate the scale-lengths of the discs.
In Table~\ref{tablescales}, we list the scale-length $h_o$ of the
old population, the scale-length $h_y$ of the young population,
and their ratio $h_o/h_y$, except for the simulations M12z and M11,
for which an exponential profile does not fit the distribution
of young stars. The scale-length is usually 
larger for the old population than the young one, with ratios varying
from 0.98 to 1.60.
This is in agreement with the recent 
observational work to \citet{yd06} if we make a direct 
correspondence between old stars and thick-disc stars.

{\it By definition\/}, by the time the merger is completed,
old stars have all formed, while the matter destined 
to form young stars is still in the form of gas. 
Hydrodynamical processes, such as oblique shock
waves, can potentially exert a torque on the gas, modifying its
angular momentum. Such processes would not affect the old population
directly, only indirectly through the gravitational interaction
between the old stars and the remaining gas. This might result in a
misalignment between the young and old discs. To investigate this
issue, we performed a least-square fit of a plane to the final distribution
of young stars, and to the final distribution of old stars. 
%For the latter,
%we only included stars with $|Z|<5\,{\rm kpc}$, in an attempt to segregate
%thick-disc stars from halo stars. 
The second column of Table~\ref{angles}
shows the angle $\theta_{oy}$
in degrees between the two planes, that is, the angle between
the young and old discs. The angles are $6^\circ$ or less
for all the runs, indicating that the discs tend to be coplanar. 
To estimate whether these angles are significant, we fitted the distribution
of stars in the old disc using an oblate ellipsoid. We then calculated
a {\it relevance angle} $\theta_o$, defined as the arc tangent of the 
short-to-long axis ratio of the ellipsoid. If we were to fit the old disc
inside a flat square box, $\theta_o$ would be the angle between the
edge and the corner, as seen from the center, so it is a measure
of how thick the old disc is. If $\theta_{oy}<\theta_o$,
then the young disc is entirely embedded inside the old disc, in spite of
the angle between them. But if $\theta_{oy}>\theta_o$, then the young
disc ``sticks out'' of the old disc. The third column of Table~\ref{angles}
shows the relevance angles. Interestingly, 
the simulation rM12, the only one for which one galaxy is
spinning in the retrograde direction, produces an old disc that is
actually over an order of magnitude thinner than the ones produced
by the other simulations 
%(this can be seen also in Fig.~\ref{rM12}).
It is well-known that a retrograde
disc will suffer very little tidal disruption prior to merger compared
to a prograde disc \citep{tt72}. This could explain why the old disc
in simulation rM12 ends-up
being quite thin, though a more detailed investigation is needed to
confirm this hypothesis.
The last column of Table~\ref{angles} shows the ratio
$\theta_{oy}/\theta_o$.
We find that all cases except rM12 have $\theta_{oy}/\theta_o<1$. Hence,
young discs tend to be embedded inside old discs. 

% From this analysis, we conclude that the structural properties
% of the remnants are strongly dependent of the initial 
% conditions.  
%Unless otherwise, major
%conclusions drawn in the subsequent sections
%are based upon the six disc-dominated remnants (that is, all but
%M11 and M12z).

\begin{table}
 \caption{Scale-length ratio for the young and old populations}
 \begin{tabular}{@{}lccc@{}}
  \hline
Run & ${h_o}$[kpc] & ${h_y}$[kpc] & $h_o/h_y$\\
  \hline
 M12    & 5.56 & 5.27     & 1.06     \\
 M1290  & 6.22 & 6.05     & 1.03     \\
 M12orb & 8.97 & 5.59     & 1.60     \\
 M12z   & 6.55 & $\ldots$ & $\ldots$ \\
 rM12   & 7.54 & 5.21     & 1.45     \\
 M11    & 5.94 & $\ldots$ & $\ldots$ \\
 M13    & 5.37 & 4.47     & 1.20     \\
 M110   & 7.71 & 7.87     & 0.98     \\
\hline
\label{tablescales}
\end{tabular}
\end{table}

\begin{table}
 \caption{Angle between the plane of the disc for 
  by the young and old populations}
 \begin{tabular}{@{}lrrr@{}}
  \hline
Run & $\theta_{oy}$[degrees] & $\theta_o$[degree] &
$\theta_{oy}/\theta_o$\\
\hline
 M12    &  3.28 & 17.62 & 0.19 \\
 M1290  &  0.70 & 11.19 & 0.06 \\
 M12orb &  3.26 & 20.64 & 0.16 \\
 M12z   &  6.00 & 41.17 & 0.78 \\
 rM12   &  2.21 &  1.27 & 1.74 \\
 M11    &  2.31 & 26.68 & 0.08 \\
 M13    &  5.63 & 17.82 & 0.32 \\
 M110   &  0.25 & 13.80 & 0.02 \\
\hline
\end{tabular}
\label{angles}
\end{table}

\subsection{Kinematics}

\begin{figure}
\begin{center}
\includegraphics[width=3.3in]{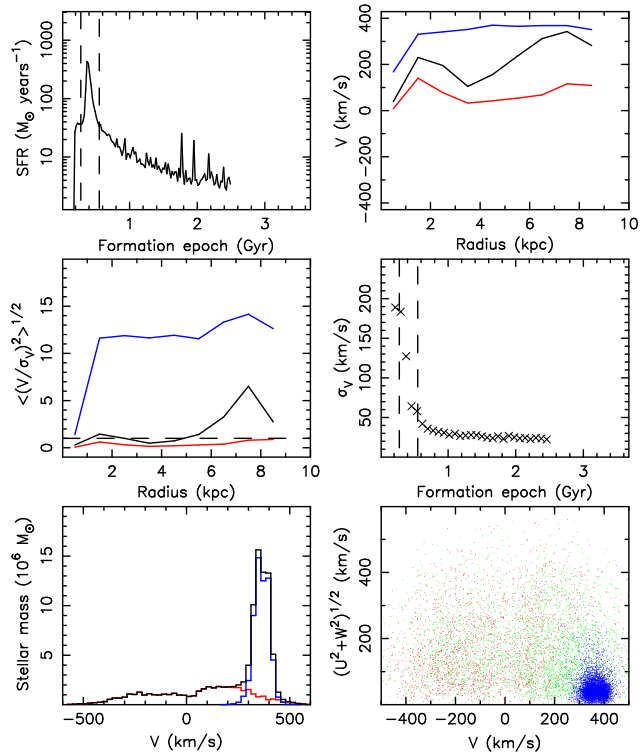}
\caption{Star formation rate and kinematics for simulation M12.
On each panel, the red curves
represent old stars, the blue curves and dots represent young stars,
and the black curves represents all stars.
The vertical dashed lines indicate the beginning and the end of the merger.
Top left: star formation rate vs. formation epoch. 
$t=0$ corresponds the the beginning of the simulation.
Top right: rotation velocity vs. radius for stars located near
the plane of the disc ($|Z|\leq1\,{\rm kpc}$). 
Middle left: Rotational support versus radius.
The dashed line separates rotationally-supported stars (above)
from stars supported by velocity dispersion (below). Middle right: velocity
dispersion vs formation epoch. Bottom left: histogram of stellar
mass vs. rotation velocity, withe negative values
indicating counter-rotating stars. Bottom right: Toomre diagram.
The black, green, and blue dots indicate stars formed before, during,
and after the merger.
}
\label{M12_kin}
\end{center}
\end{figure}

%\begin{figure}
%\begin{center}
%%\includegraphics[width=3.3in]{KinM12orbB.ps}
%\includegraphics[width=3.3in]{figure12.eps}
%\caption{Same as Fig.~\ref{M12_kin}, for simulation M12orb.}
%\label{M12orb_kin}
%\end{center}
%\end{figure}

\begin{figure}
\begin{center}
\includegraphics[width=3.3in]{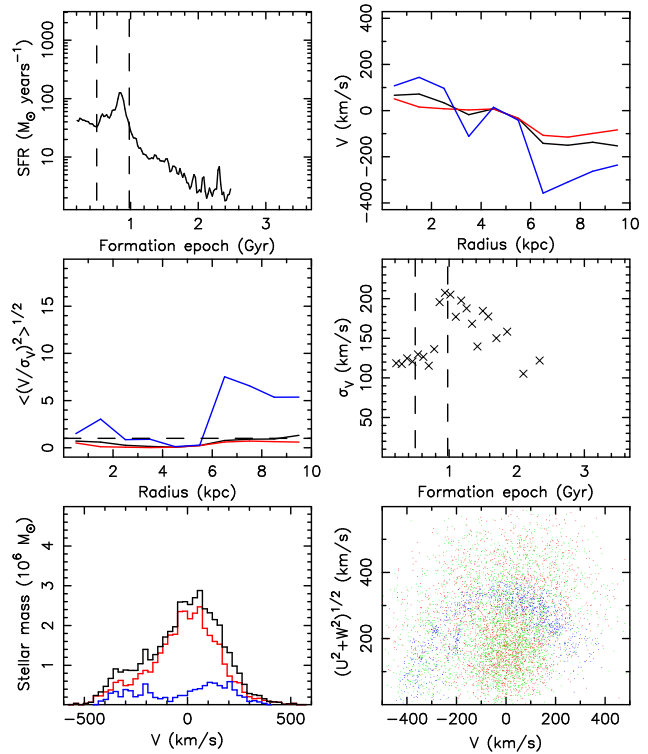}
\caption{Same as Fig.~\ref{M12_kin}, for simulation M12z.}
\label{M12z_kin}
\end{center}
\end{figure}

%\begin{figure}
%\begin{center}
%%\includegraphics[width=3.3in]{KinM1290B.ps}
%\includegraphics[width=3.3in]{figure14.eps}
%\caption{Same as Fig.~\ref{M12_kin}--\ref{M12z_kin}, for simulation M1290.}
%\label{M1290_kin}
%\end{center}
%\end{figure}

\begin{figure}
\begin{center}
\includegraphics[width=3.3in]{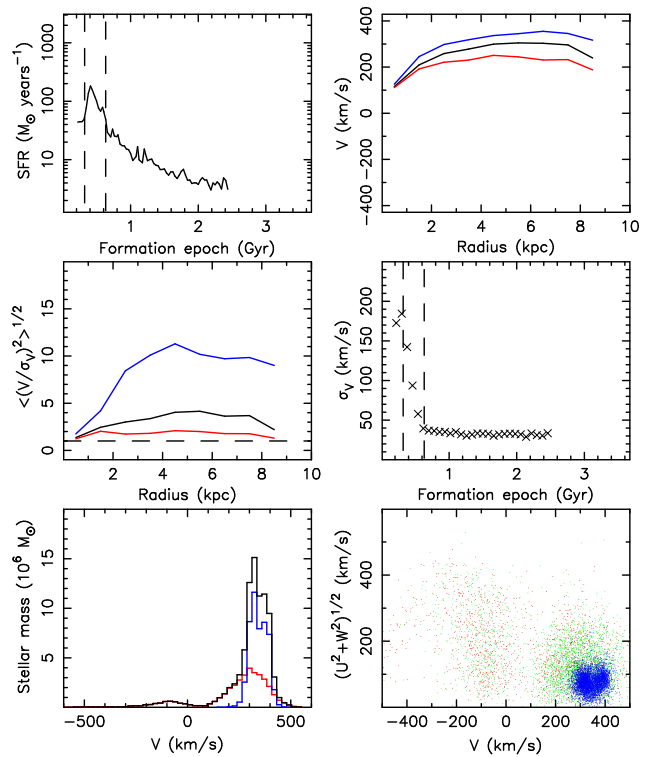}
\caption{Same as Fig.~\ref{M12_kin}--\ref{M12z_kin}, for simulation rM12.}
\label{rM12_kin}
\end{center}
\end{figure}

\begin{figure}
\begin{center}
\includegraphics[width=3.3in]{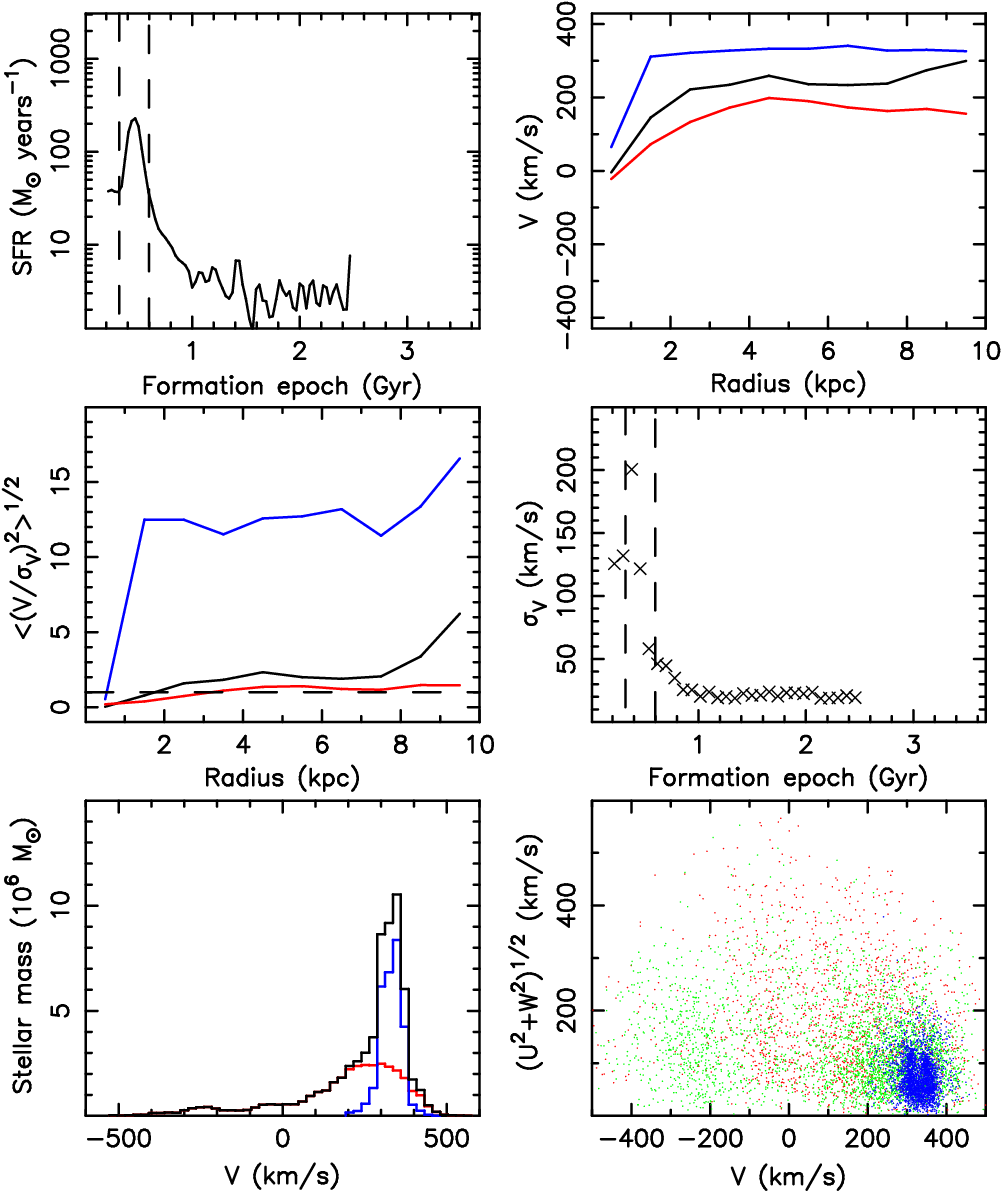}
\caption{Same as Fig.~\ref{M12_kin}--\ref{rM12_kin}, for simulation M13.}
\label{M13_kin}
\end{center}
\end{figure}

We have analyzed the kinematic properties of all 8
remnants. 
We computed the rotation curves of the old and young
populations, for all the remnants. The top right panels
of Figs.~\ref{M12_kin}--\ref{M13_kin} show some examples of
these curves. 
For simulations M12, M12orb, M1290, rM12, M13, and M110, the old 
populations have a lower mean rotational velocity than their 
younger counterparts. The difference is about $200-300\,{\rm km\,s^{-1}}$
for M12, M12orb, and M1290 (mass ratio 2:1), $150\,{\rm km\,s^{-1}}$
for M13 (mass ratio 3:1) and $50\,{\rm km\,s^{-1}}$ for M110
(mass ratio 10:1).
Hence, it decreases with increasing mass ratio of the progenitors.
Comparing M12 with rM12, we also find that
reversing the rotation of Gal2 (so that it is retrograde relative to
the orbit of the galaxies) also reduces the differences between the
two populations (about $300\,{\rm km\,s^{-1}}$ for M12 versus
$150\,{\rm km\,s^{-1}}$ for rM12).
The rotation curve of M12z and M11 are drastically different.
The young population of the outer ring is counter-rotating with respect
to the young population of the outer disc.
We also found some counter-rotating stars
in other simulations, but (i) they belong to the old population,
and (ii) these stars were in the minority, so the overall rotation curve
showed no counter-rotation.

%\begin{figure}
%\begin{center}
%%\includegraphics[width=3.3in]{KinM11B.ps}
%\includegraphics[width=3.3in]{figure16.eps}
%\caption{Same as Fig.~\ref{M12_kin}--\ref{rM12_kin}, for simulation M11.}
%\label{M11_kin}
%\end{center}
%\end{figure}

%\begin{figure}
%\begin{center}
%%\includegraphics[width=3.3in]{KinM110B.ps}
%\includegraphics[width=3.3in]{figure18.eps}
%\caption{Same as Fig.~\ref{M12_kin}--\ref{M13_kin}, for simulation M110.}
%\label{M110_kin}
%\end{center}
%\end{figure}

To estimate the relative importance
of rotation and velocity dispersion in providing support against gravity,
we calculated for each population the average rotational velocity $V$ 
and its dispersion $\sigma_V^{\phantom2}$ in radial bins.
The middle left panels
of Figs.~\ref{M12_kin}--\ref{M13_kin} the quantity 
$\langle(V/\sigma^{\phantom2}_V)^2\rangle^{1/2}$ versus radius,
for some of the remnants. 
In all simulations, we find $\langle(V/\sigma^{\phantom2}_V)^2\rangle^{1/2}>1$
for the young population, indicating that this population forms a
massive, rotationally supported disc. 
The only exception is
simulation M12z, where $\langle(V/\sigma^{\phantom2}_V)^2\rangle^{1/2}<1$
between radii 3 and $6\,\rm kpc$. 
As the top panel of Fig.~\ref{M12z_kin} shows, this region corresponds to the
discontinuity between the inner disc and the
counter-rotating outer ring.
Simulations M12, M12orb, M12z, and M11 all have 
$\langle(V/\sigma^{\phantom2}_V)^2\rangle^{1/2}<1$, indicating the presence
of a ``hot disc'' supported by internal motions rather than rotation.
Simulation M13 has $\langle(V/\sigma^{\phantom2}_V)^2\rangle^{1/2}<1$
at radii $r<3\,{\rm kpc}$ and 
$\langle(V/\sigma^{\phantom2}_V)^2\rangle^{1/2}>1$ at larger radii.
Simulations M1290, rM12, and M110 all have
$\langle(V/\sigma^{\phantom2}_V)^2\rangle^{1/2}>1$, indicating that
the old population is also rotationally supported, but in all three cases
the rotational support is significantly smaller for the old population,
typically by a factor of order 5.
In the case of simulation rM12, this result is consistent
with the fact that retrograde orbits are less destructive than prograde
orbits.

The middle right panels of Figs.~\ref{M12_kin}--\ref{M13_kin} show
the velocity dispersion versus formation epoch, with the dash vertical
lines indicating the starburst. In 5 of the 8 simulations, the velocity
dispersion is larger for stars formed prior to the collision, and smaller
for stars born after the collision. This is consistent with the statement
that old stars have a smaller rotational support than young stars. The
exceptions are the simulations with high mass ratios, M13 and M110, for
which $\sigma^{\phantom2}_V$ peaks during the starburst, and M12z,
for which $\sigma^{\phantom2}_V$ peaks after the starburst 
(Fig.~\ref{M12z_kin}).

The bottom left panels of Figs.~\ref{M12_kin}--\ref{M13_kin} show
histograms of the stellar mass versus rotation velocity.
For simulations M12, M12orb, M1290, rM12, and M13,
the young
population is concentrated in a narrow region of the histogram,
while the old population is spread in velocity, from the
positive to the negative side (though it is concentrated mostly
on the positive side for rM12).
The young component is counter-rotating for simulation M11.
For simulations M12z, parts of the young component are
counter-rotating, and the total distribution is centered near
zero. Simulation M110 (the minor merger) differs from all others
in that the old population is concentrated at positive velocities.
Notice that this is the only case for which the old population is
rotationally supported.
% (middle left panel of Fig.~\ref{M110_kin}).

One way to discriminate between kinematically different star populations 
is the use of the Toomre Diagram \citep{sf87}. By comparing $V$ (circular 
velocity) with $U+W$ (radial and perpendicular velocity respectively), we can 
have a clearer picture of the kinematic distribution of the young and old 
population. A thin disc will be mostly formed by stars with low $U+W$ and a 
large $V$ component since all stars will be in nearly coplanar orbits.
The bottom right panels of Figs.~\ref{M12_kin}--\ref{M13_kin} show the 
Toomre diagram for each simulations.
For M12, the young stars are concentrated in a small region of
the diagram, centered around $V=400\,{\rm km\,s^{-1}}$ and
$(U^2+W^2)^{1/2}=0\,{\rm km\,s^{-1}}$, which is characteristic of a thin
disc since these stars are mainly on fast rotating planar orbits.
 The old stars are distributed throughout the
diagram, indicating that these stars follow orbits with significant radial
and orthogonal velocities, characteristic of
systems supported by velocity dispersion. These stars are located
mostly in the thick disc and halo.  

The Toomre diagrams for the other simulations show similarities,
and also interesting differences, with the diagram for simulation M12.
%The simulation with a large interaction angle (M1290,
%see Fig.~\ref{M1290_kin}) 
%produces a thick disc that is less populated than the simulations with a 
%lower angle. 
%A large angle between the two interacting disc leads to the formation of a 
%bigger halo, since the gas particles provided by the second galaxy will 
%start up with
%a larger velocity component in the $Z$-direction. 
%A larger halo necessarily implies a 
%smaller thick disc, since they are made of the same population.
Some simulations present a discontinuity between the positive and 
negative region of the diagram. For example, simulation rM12 produces
two different 
population of old stars, one forming a thick disc and another one 
forming a counter-rotating thick disc/halo. Simulation M12z and M11 
have a different Toomre diagram than the other simulations. 
For M12z (Fig.~\ref{M12z_kin}),
there is no clear disc formed in the remnant, and the young stars 
are dispersed throughout the diagram. 
As we saw in the previous section, young stars in this simulation
are mostly located in a massive ring around the galaxy.
For simulation M11, the young stars in
the outer ring are counter-rotating while all stars in the inner 
disc are co-rotating. For simulation M110, the young
population has a significant $(U^2+W^2)^{1/2}$ component
($\sim150\,{\rm km\,s^{-1}}$),
almost as large as the $V$ component.

\begin{figure}
\begin{center}
\includegraphics[width=3.3in]{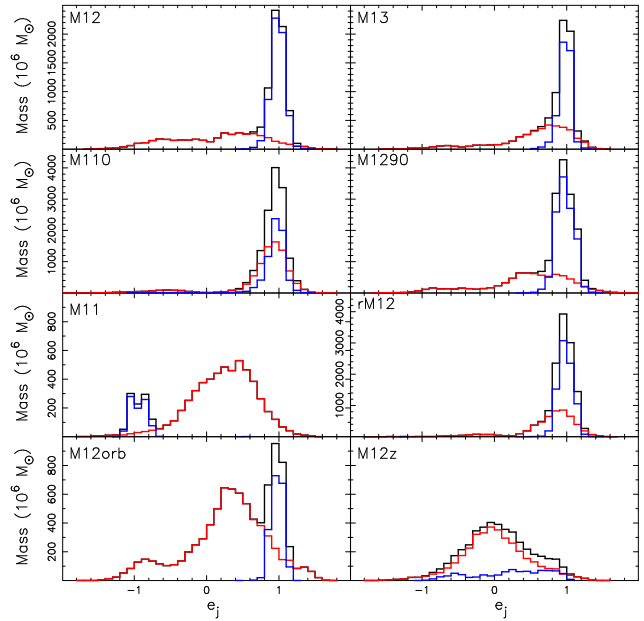}
\caption{Circularity of the orbits for all simulations.
$e_j$ represent the angular momentum of a star particle normalized to a 
particle in a circular orbit of the same radius. On each panel, the red curves
represent old stars, the blue curves represent young stars,
and the black curves represent all stars.
}
\label{e_j}
\end{center}
\end{figure}

We have also performed an analysis of the circularity of the stellar orbits,
where $e_{\rm j}=j_{\rm z}/j_{\rm circ}$, and $j_{\rm z}$ is the $z$-component
of the specific angular momentum of each star, and $j_{\rm circ}$ is the 
angular
momentum expected for a circular orbit,
following the method of \cite{pat09}, which is an approximation of a method 
used by \cite{abadietal03}. These results are showed in Fig.~\ref{e_j}. 
Simulations M12, M13, M110, M1290, rM12, and M12orb all present a young star 
population with circular orbits ($e_j\sim1.0$) and an old population with a 
high dispersion of circularity. One exception is M110, which presents near 
circular orbits for both the young and old population, which is expected
in a minor merger. Simulations M11 and M12z show 
non-circular orbits, in good agreement with the other kinematical 
results. The small thin disc of M11 is also well represented at 
$e_j\sim-1.0$.

The main conclusion we can draw from the analysis and comparison of
the kinematical properties of the remnants is that 
these kinematical properties, as the structural ones,
are strongly dependent of the initial 
conditions, and support previous results by \citet{robertsonetal06}.

\subsection{Chemical Abundances}

\begin{figure}
\begin{center}
\includegraphics[width=3.3in]{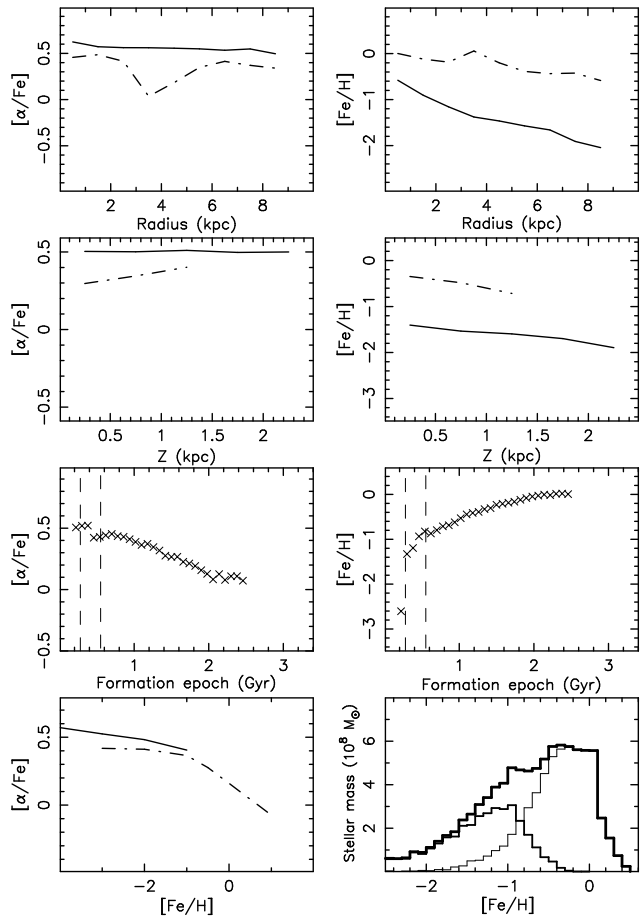}
\caption{Chemical abundances for simulation M12.
On the top four panels and bottom left panel, solid curves
represent old stars, and
dot-dashed curves represent young stars.
The first three rows
show the $\alpha$-elements abundance (left) and metallicity (right)
versus radius (first row), height (second row), and formation epoch
(third row). Bottom left panel: $\alpha$-element abundance versus metallicity.
Bottom right panel: Stellar mass versus metallicity, for stars born before
or after the merger (medium line), after the merger (thin line), and for
all stars (thick line).
}
\label{M12_chem}
\end{center}
\end{figure}

\begin{figure}
\begin{center}
\includegraphics[width=3.3in]{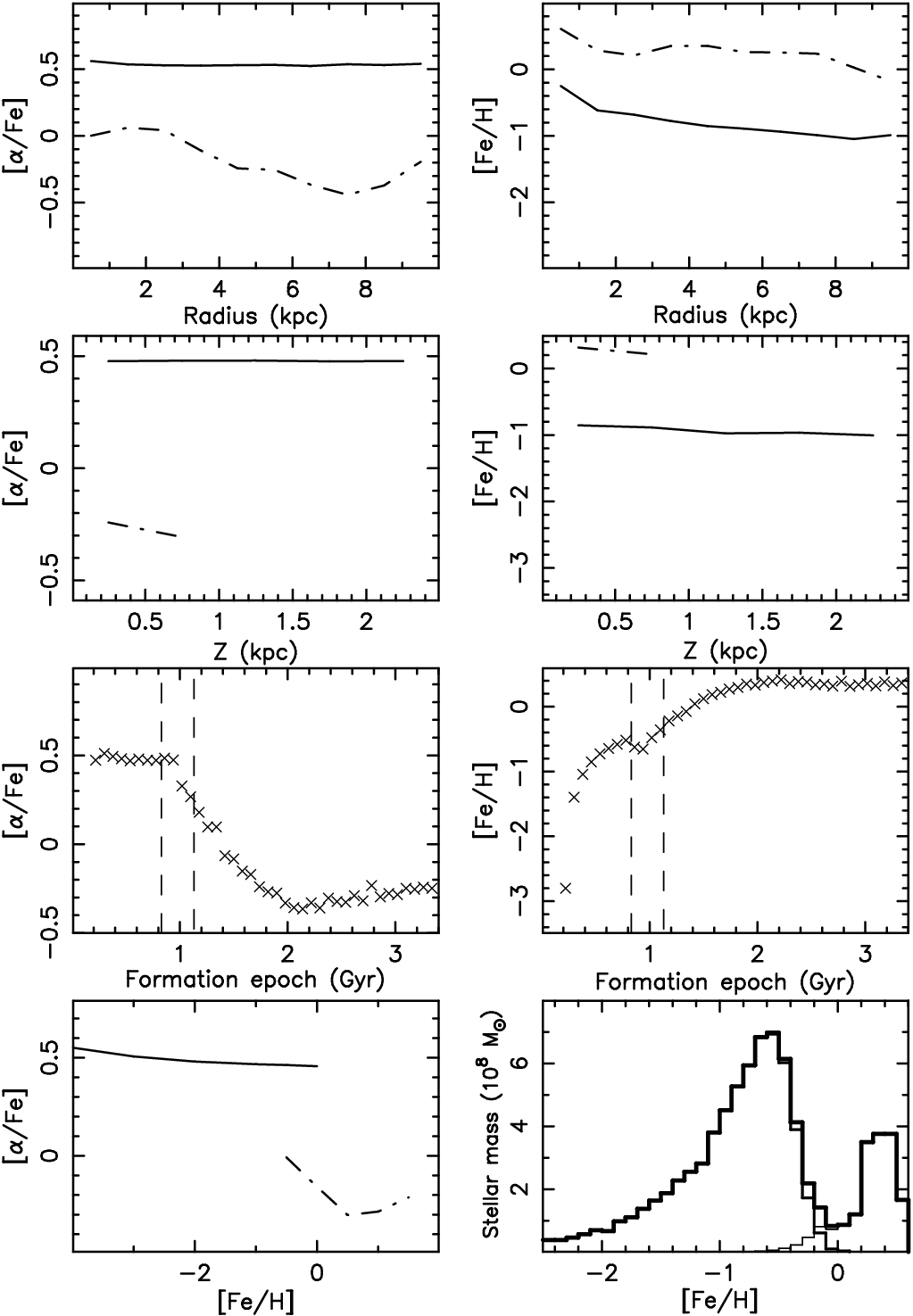}
\caption{Same as Fig.~\ref{M12_chem}, for simulation M12orb}
\label{M12orb_chem}
\end{center}
\end{figure}

\begin{figure}
\begin{center}
\includegraphics[width=3.3in]{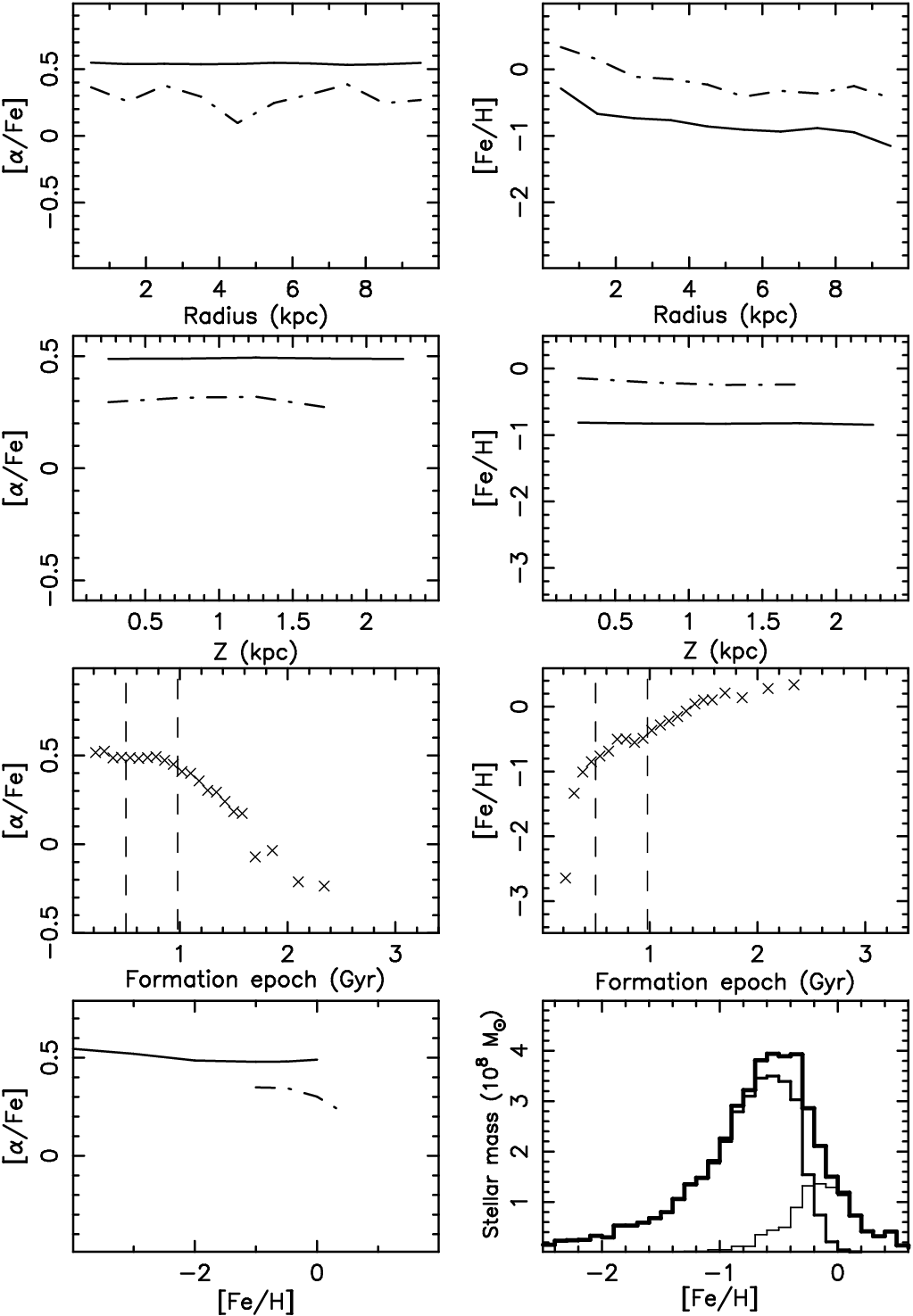}
\caption{Same as Fig.~\ref{M12_chem}--\ref{M12orb_chem}, for simulation M12z}
\label{M12z_chem}
\end{center}
\end{figure}

\begin{figure}
\begin{center}
\includegraphics[width=3.3in]{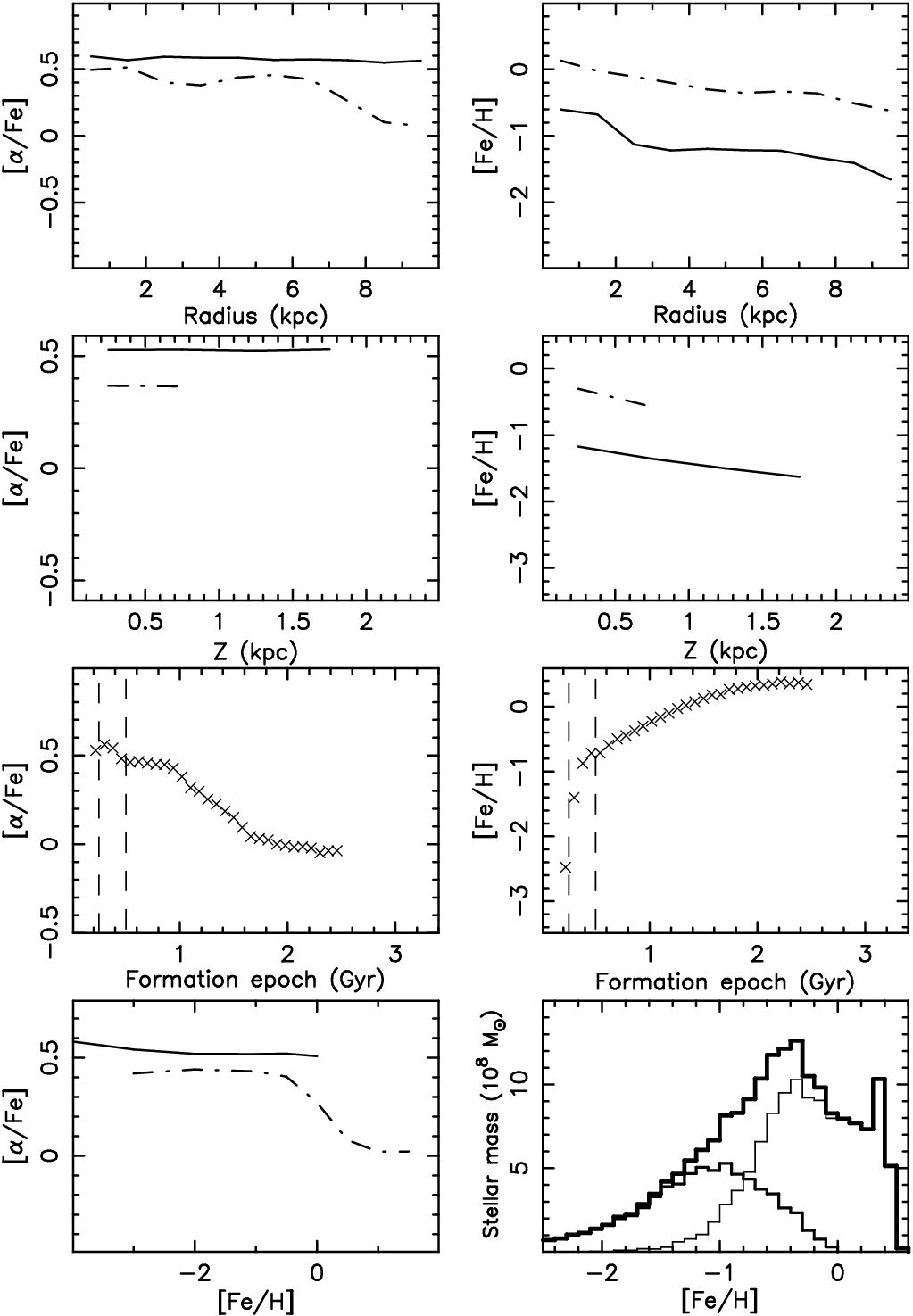}
\caption{Same as Fig.~\ref{M12_chem}--\ref{M12z_chem}, for simulation M1290.}
\label{M1290_chem}
\end{center}
\end{figure}

\begin{figure}
\begin{center}
\includegraphics[width=3.3in]{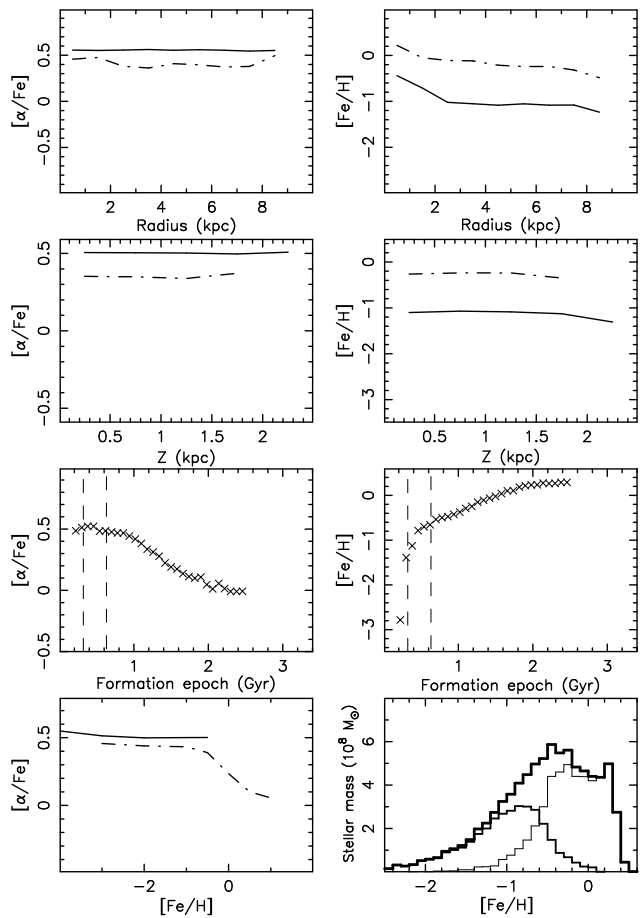}
\caption{Same as Fig.~\ref{M12_chem}--\ref{M1290_chem}, for simulation rM12.}
\label{rM12_chem}
\end{center}
\end{figure}

\begin{figure}
\begin{center}
\includegraphics[width=3.3in]{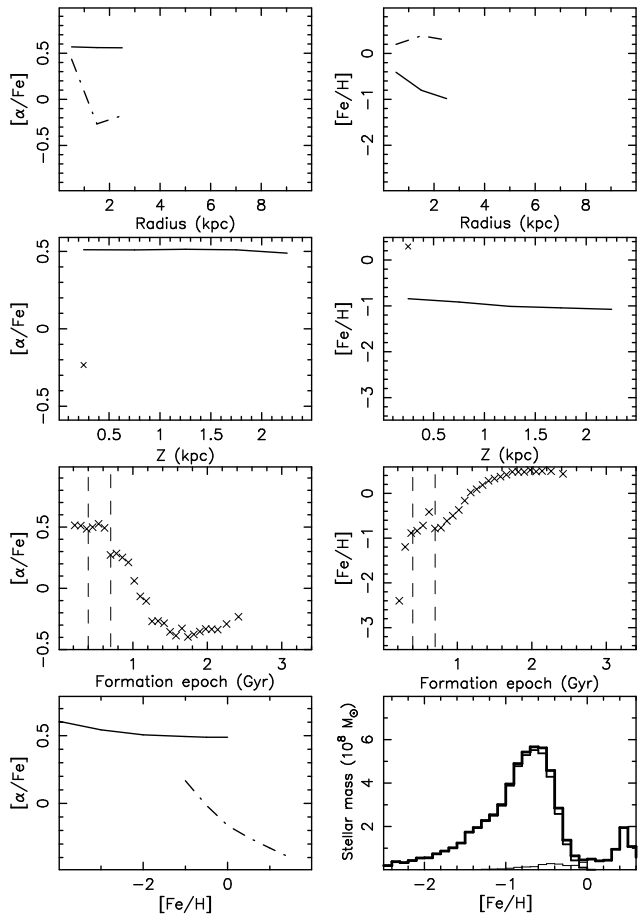}
\caption{Same as Fig.~\ref{M12_chem}--\ref{rM12_chem}, for simulation M11.
We could not calculate vertical profiles for the young stars, because the
thin disc is too thin. Instead we indicate the mean values
of \al and \fe with a single symbol ($\times$) in the second-row
panels.}
\label{M11_chem}
\end{center}
\end{figure}

\begin{figure}
\begin{center}
\includegraphics[width=3.3in]{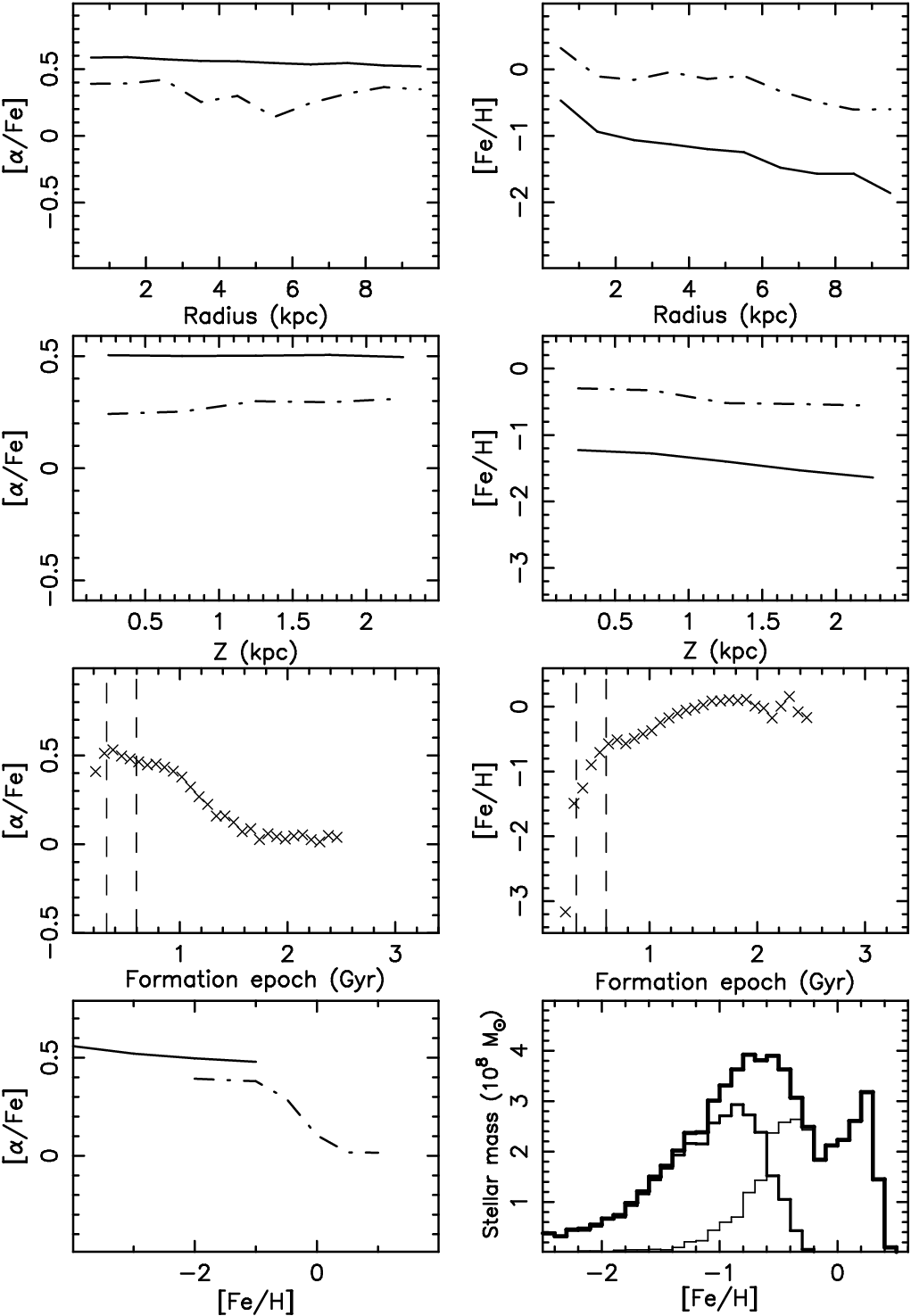}
\caption{Same as Fig.~\ref{M12_chem}--\ref{M11_chem}, for simulation M13}
\label{M13_chem}
\end{center}
\end{figure}

\begin{figure}
\begin{center}
\includegraphics[width=3.3in]{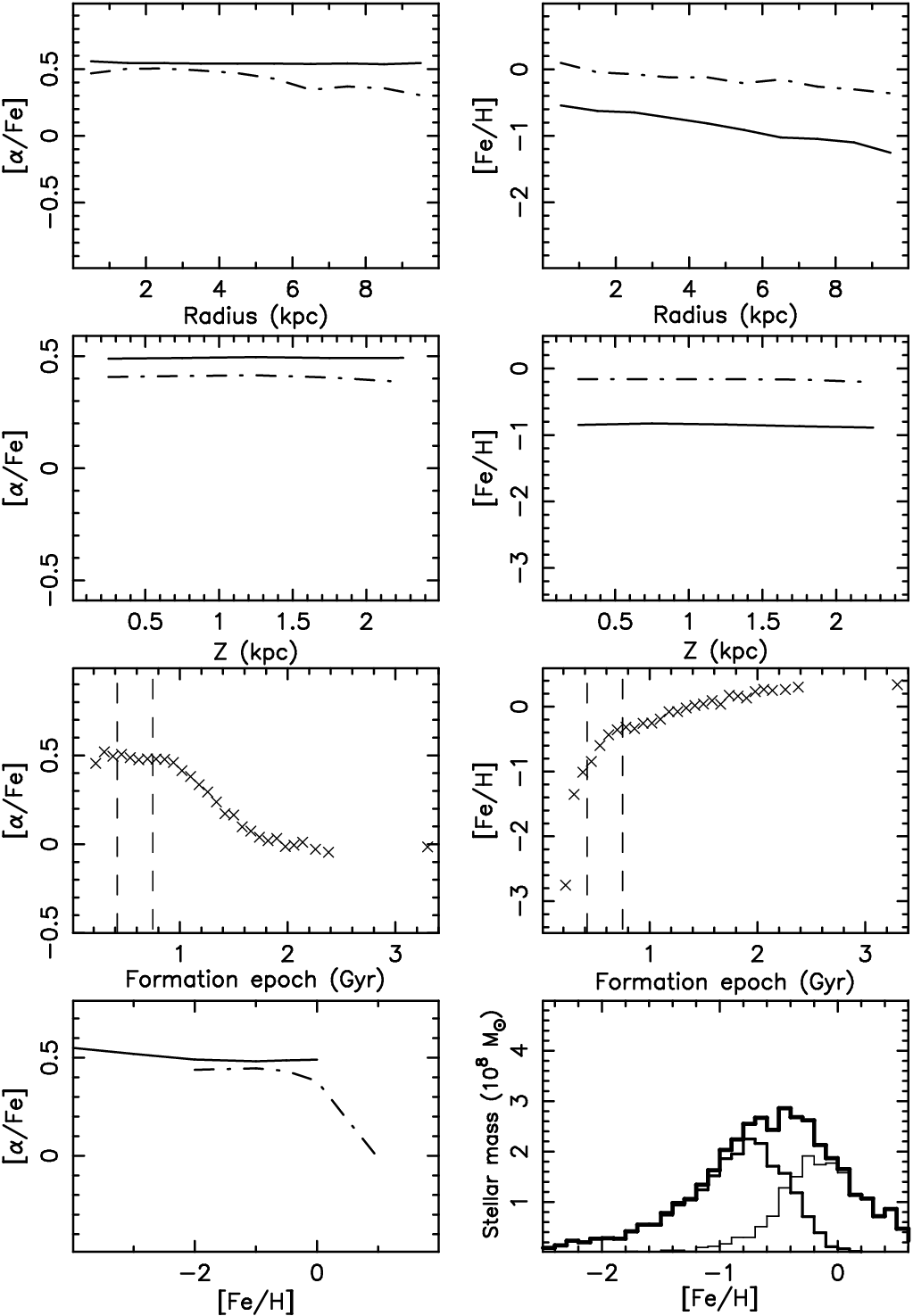}
\caption{Same as Fig.~\ref{M12_chem}--\ref{M13_chem}, for simulation M110}
\label{M110_chem}
\end{center}
\end{figure}

As mentioned in the Section~I, one of the most important results not 
yet achieved concerning gas-rich major mergers is their chemical signature. 
Here we present some of our first results. 
The chemical properties of all remnants are presented in 
Figs.~\ref{M12_chem}--\ref{M110_chem}. The top panels show
the radial profiles of the $\alpha$-element abundances \al
(calculated by averaging the abundances of O, Mg, and Si)
and metallicity \fe, calculated by averaging over radial bins. 
The old population has a higher \al ratio and lower
metallicity \fe than the young population, at all radii. 

The gradients of \al in the old population are all very flat. 
The distributions of
\al in the young population are much more complex. In particular,
simulations M12, M12z, M1290, and M13 show several local minima and
maxima at various radii, with variations as large as 0.5 dex.
The gradients of \fe are much smoother.
In all simulations except M11, we find a gradient in 
$\rm[Fe/H]$, ranging from $-0.01$ to $-0.1$~dex per kpc.

\begin{figure*}
\begin{center}
\includegraphics[width=4.8in]{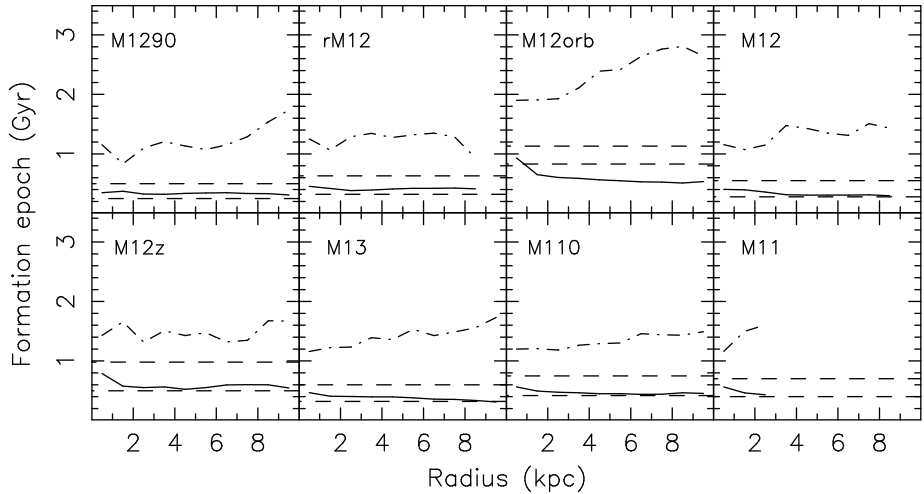}
\caption{Star formation epoch versus final radius for old stars (solid line)
and young stars (dot-dashed line), for the 8 simulations. The horizontal
dashed lines indicate the beginning and the end of the starburst.}
\label{ageradM12}
\end{center}
\end{figure*}

The third rows of panels
in Figs.~\ref{M12_chem}--\ref{M110_chem}, show the \al and \fe ratio
versus formation epoch. The \al ratio remains fairly 
constant over time during the merger, and starts decreasing
after the merger.
The \fe ratio shows a completely opposite behavior: it increases
strongly before and during the merger, and keeps increasing, but more
slowly, after the merger. This 
would suggest that the galaxy grows outside-in, with the
older, metal-poor stars being located at the outer regions of the disc.
To investigate this question, we plot in
Fig.~\ref{ageradM12} the formation epoch of old and young
stars versus their final location in the disc, for all simulations.
The dashed lines indicate the beginning and the end of the merger.
For old stars, the formation epoch either decreases with increasing radius
(as in M12) or remains constant (as in M1290).
For instance, in the case of M12, stars that formed before the merger
 are dominant at radius R$>$3 kpc, while stars that formed during the merger 
 and the associated starburst are more centrally concentrated.
% For instance, in the case of M12,
% stars that formed before the merger are located at radius $R>3\,{\rm kpc}$,
% while stars that formed during the merger and the associated starburst
% are centrally located at radius $R<3\,{\rm kpc}$. 
The young stars show the
opposite trend: the formation epoch increases with radius up to
$R=7.5\,{\rm kpc}$ in most simulations. 
Star-formation after the merger therefore proceeded inside-out.
It is the efficiency of chemical enrichment, and not the epoch of
star formation, that explains the gradient in $\rm[Fe/H]$. 
Chemical enrichment is more efficient in the center of 
the remnant, where the stellar density is higher. 

The second row of panels
in Figs.~\ref{M12_chem}--\ref{M110_chem} show \al and \fe versus
height above the plane. The different extents of the two curves
on each panel reflects the different thicknesses of the two discs.
Notice that the thin disc is {\it very thin\/} in simulation M11,
less than 800~pc.
There is essentially no vertical gradient in \al
for the old population. For the young population, 
we find a gradient in \al for M12 (0.1 dex per kpc), 
and possibly another one for
M12orb, while there are no evidence for gradients in the
other cases.
The \fe ratios are
either constant, as in simulations M12z and M110, or drops by
a factor of order 2 or 3, as in simulations M12 and M1290,
for both the old and the young populations.
Since the stellar density, and the resulting chemical enrichment,
is a much stronger function of radius than height, we naturally expect
the vertical gradient of the metallicity to be much weaker than the radial 
one \citep{brooketal05}.

The most important plots on Figs.~\ref{M12_chem}--\ref{M110_chem} 
are the \al versus \fe plots (bottom-left panels). Old stars have larger 
$\alpha$-element abundance than young stars, up to a relatively high
metallicity ($\rm[Fe/H]\simeq-0.5$). 
%This motivated us to publish
%the results of our initial simulation, M12, in an earlier paper
%(B07). We now find the same result in all simulations,
%including the two simulation M12z and M11 where a disc 
%structure is not obvious. 
We can explain this result by considering
the relations between $\rm[\alpha/Fe]$, $\rm[Fe/H]$,
and the formation epoch of stars (third-row
panels in Fig.~\ref{M12_chem}--\ref{M110_chem}). The metallicity \fe increases
with time during the starburst, and levels-off after the starburst is
completed.
The ratio \al then decreases with
time as the metallicity increases, until it reaches the values found in the 
thin disc. This decrease of \al is slowed near the beginning of the 
collision, as the starburst leads to a large number of Type~II SNe,
that enrich the gas in $\alpha$-elements. After the starburst, Type~Ia
SNe become effective, and enrich the gas in iron.
%, while intermediate-mass stars produce $\alpha$-elements. 
This explains why the ratio \al
gradually decreases after the collision. 

\begin{figure*}
\begin{center}
\includegraphics[width=4.8in]{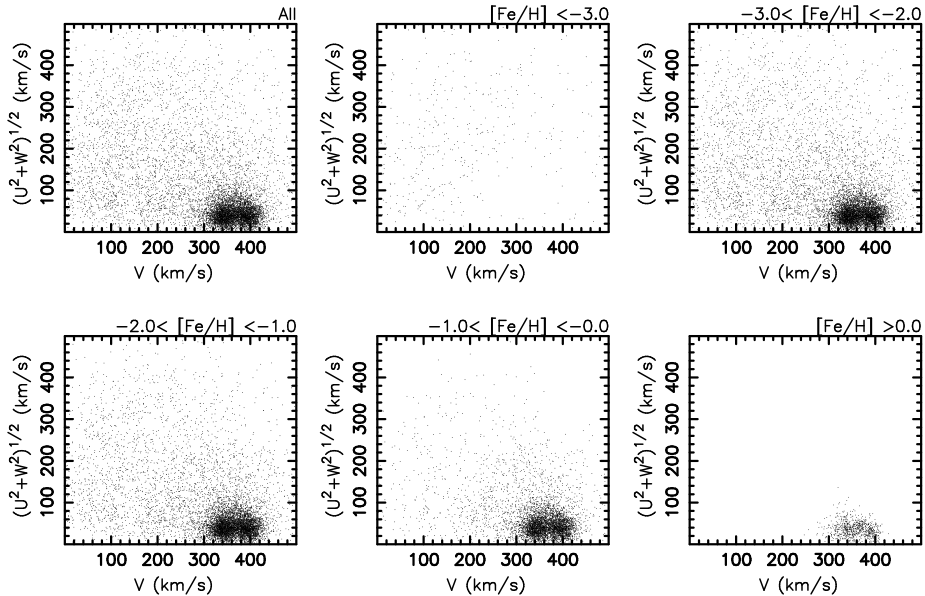}
\caption{Toomre diagram for stars located near the
disc plane ($Z<1\,{\rm kpc}$), for simulation M12,
divided according to metallicity.
The metallicity range is indicated on top of each panel.}
\label{toomreM12}
\end{center}
\end{figure*}

We complete this study by calculating the stellar mass in metallicity bins, for
both populations. The results
are shown in the bottom right panels of 
Figs.~\ref{M12_chem}--\ref{M110_chem}.
Old and young stars have different metallicity distributions. The old 
population peaks in the range $\rm[Fe/H]=[-1.0,-0.6]$, 
while the young population peaks in the range
$\rm[Fe/H]=[-0.4,0.4]$. The implies that old stars are formed, on average,
by gas that has not yet been enriched by a large number of nearby
Type~Ia SNe. We can gain more insight into this process by plotting Toomre
diagrams for different metallicity bins. 
The results for simulation
M12 are shown in Fig.~\ref{toomreM12}. We recall that the concentration
of stars in the region $V\simeq400\rm\,km/s$, 
$(U^2+W^2)^{1/2}\simeq0\rm\,km/s$ represents the thin disc, and the reminder
of the diagram represents the thick disc and the halo.
Low-metallicity stars 
($\rm[Fe/H]<-3$) are located in the halo, high-metallicity stars 
($\rm[Fe/H]>0$) are located in the thin disc, and intermediate-metallicity
stars are found in both locations.

The main result we draw from this analysis of the chemical abundance is
that the merger leaves a clear signature: the old stars, located in the
thick disc and the halo, have a ratio \al that remains constant 
with increasing $\rm[Fe/H]$ up to high metallicities ($\rm[Fe/H]=-0.5$),
while the young stars, located in the thin disc, have a lower
ratio \al which decreases with increasing metallicities. 
This result is very robust: we found it in each of the
8 simulations included in our study, including the minor merger
(simulation M110). 

Early gas-rich mergers in our simulations occur, by construction, 
prior to the timescale for significant Type~Ia SNe pollution.
Ultimately, this is responsible for the robustness of our results.

\section{DISCUSSION}

The results presented 
in this paper support the gas-rich merger scenario as 
a way to explain the formation of large disc galaxies in the Universe. 
Our results 
agree very well with the kinematical results of \citet{robertsonetal06}.
%and the fact that gas fraction
%in the progenitors must be in excess of 50\% prior to merging. 
That said, the gas-rich process
may be necessary to form a disc, but it is not a sufficient
condition. We can see this in our two simulations which formed
bulge-dominated galaxies, namely M11 and M12z. 
Explaining this difference is not simple because of 
the chaotic nature of the merger process. There is 
likely a complex relationship between the initial conditions and the
behavior of the gas during the merger. 
The duration of the merger and the violence of 
the collision seems to favor a higher dynamical warming, and this favors 
the formation of a small rotationally supported disc, embedded within a 
large spheroidal component.
In the case of simulation M11, the two progenitors 
have a lower mass than in the others simulations, but the 
starburst has the same 
intensity. As a result, there is not enough gas left after the merger
to form a substantial thin disc. 
This agrees with the results of \citet{robertsonetal06} 
who found that low-mass progenitors are more likely to form an 
important spheroidal component than high-mass ones. 

When a disc-like remnant is formed, it invariably consists of
two disc components: a thick disc 
made of stars formed before or during the merger, and a thin disc made
of stars born after the merger. The thick disc has a lower rotation velocity 
than the thin disc, in  agreement with the findings of \citet{yd06}. 
The scale-length ratio between the two components is also in agreement 
with this work, which suggests that gas-rich disc mergers may be a common
way to form disc galaxies. However, we have to be careful since subsequent gas 
infall from satellites or the intergalactic medium is not included in 
our simulations. Such 
gas infall could increase the scale-length of the thin disc and affect
our results (see, e.g., \citealt{brooksetal09}). 
Stars belonging to the thick and thin discs differ by their abundances
in various metals. Old stars have an excess in $\alpha$-elements relative
to young stars at metallicities \fe below $-0.5$.
This result is in good agreement with 
observational studies on the chemical abundances of the two disc components 
of the Milky Way,
if the old and young stars in our simulations
are analogous to the thick- and thin-disc stars, 
respectively. \citet{rlap06} published a spectroscopic 
survey of 176 stars with high probability to be part of the galactic thick 
disc, selected with the {\sl Hipparcos} catalog. 
These authors analyze the relationship between the abundances of 22 
chemical elements and the \fe ratio. It is clear form their
Fig.~12 that the \al ratio is higher for the thick disc 
than the thin disc by almost 0.15 until the metallicity reaches
$\rm[Fe/H]=-0.3$. At larger metallicities, the \al ratios of the two discs
becomes similar. The authors explain this fact by invoking an extended period
of enrichment in iron by Type Ia SNe, an explanation that is supported
by our work.
Fig. 16 in the same article shows that chemical elements other than the 
$\alpha$-elements do not follow the same pattern, which is further 
evidence of the important role of Type II SNe in the formation of 
the Milky Way, and support the scenario that a major collision took place
in the history of the Milky Way.

Another similar study was done by
\citet{bensbyetal05}. That study presents abundances for 102 dwarf 
F and G stars, and uses their kinematics to determine if they belong
to the thin or thick disc. Their Fig.~8 shows the abundances of oxygen, 
magnesium, and silicon as functions of metallicity,
which indicates that
that stars in the thick and thin disc have different abundances for
a metallicity $\rm[Fe/H]<-0.5$. Their Fig. 10 is particularly interesting
and shows that the abundance of oxygen decreases with increasing metallicity
in the interval $\rm-1.0<[Fe/H]<0.0$. This is all
in agreement with the results
presented here. These authors suggest the existence of several
observational constraints related to the formation and chemical evolution
of the thick disc: 

\begin{itemize}

\item Thick-disc stars and thin-disc stars have different chemical 
abundances.

\item For a \fe smaller than a certain value, thick disc stars have
a larger abundance of $\alpha$-elements than thin-disc stars.

\item The ratio \al decreases with increasing $\rm[Fe/H]$,
which shows the important contribution of Type~Ia SNe.

\item Stars in the thick disc are on average older than stars in the thin 
disc.

\end{itemize}

Our results mostly agree with these constraints, which supports the
gas-rich, major-collision scenario for the formation of the thick disc
of the Milky Way, a scenario in which a collision leads to an intense starburst
accompanied by a very large number of Type~II SNe. These SNe enrich
the gas supply in $\alpha$-elements, allowing the formation of
stars with a high
\al ratio. The starburst also leads to a progressive enrichment in iron caused
by Type~Ia SNe. Stars formed after the collision will therefore have
a smaller \al for the same metallicity. These stars will form a thin disc with 
a smaller velocity dispersion than the stars formed during the merger. Another
interesting result is the drop in \fe with increasing radius. 

Finally, we must point out that the initial conditions used in
our study are greatly simplified. First, the initial conditions do not
include a spheroidal component made of stars. We assume that the merger 
progenitors are pure disc for simplicity. We might expect that these
stars, if included, would end up in the halo of the final galaxy, which
might explain that most simulations produced a low-mass halo.
We can argue that our results are valid in the limit where the
initial spheroidal components are too small to affect the dynamics of
the collision. We have not included a bulge component as well.
Although morphologies of high redshift galaxies are unknown,
this assumption may be too simplistic and our results may be
biased by this assumption. However, a recent work showed the presence of a 
bulge can produce a disc remnant with a gas ratio as low as 12\% 
\citep{barnes02}.
These facts are an evidence that a merger between bulgeless disc-disc galaxies
is not a sufficient condition to form a dick remnant.
In fact, bulgeless mergers generally produce elliptical remnants
 in dissipationless simulations \citep{GB05}.
So, the presence of a massive gaseous component is probably the major condition
to reform a disc after a major merger.
We also have to add that the softening length used in this study is 
slightly higher than what are used
in the recent state-of-the-art simulations. However, the morphological
aspects of our study show good agreement with previous such higher spatial
resolution simulations, showing that the resolution is sufficient
for the analysis presented in this paper.

The hierarchical model of structure formation normally includes a significant
amount of collision and accretion of low-mass galaxies. It is presumptuous
to assume that the formation of the thick disc can be explained entirely by
one single collision. It is possible that several of these collisions
take place in the initial phases of the formation of spiral
galaxies \citep{brooketal05,conselice06}. Nevertheless, our study
can explain several observations with the important starburst that takes place 
during a major, gas-rich collision.

\section{SUMMARY AND CONCLUSION}

Using GCD+, we have performed 8 simulations of major mergers between gas-rich
spiral galaxies. We have
analyzed the kinematic, structural, and chemical properties of stars formed
before and during the collision (the old population) and stars formed after 
the collision (the young population). We used the
star formation rate to define these two populations.
A fraction of the old stars end up in the halo of the merger remnant,
while the remaining stars form a thick disc which is partly
supported by velocity dispersion, partly supported by rotation, and
sometimes includes a significant counter-rotating component. 
The young stars form a thin disc that is supported by rotation,
and, in many cases, a ring that might or might not be coplanar with
the thin disc. The disc themselves tend to be coplanar, the angle between
them being $\sim6^\circ$ or less.
With rare exceptions, both discs are well-fitted by
an exponential profile, and the scale-length of the thick disc
exceeds the one of the thin disc a few percents up to
a factor of 1.60.

The starburst occurring during the collision rapidly enriches the gas in
various metals. Explosions of Type~II SNe follow rapidly the start of the
collision, owing to the short lifetime of their progenitors, and enrich
the intergalactic medium in $\alpha$-elements. Enrichment by Type~Ia SNe
is spread over a large period of time,
enabling the enrichment in iron of both stellar
populations, which results in an old population having a \al ratio higher
than the young population, even at relatively large metallicities
($\hbox{\fe}=-0.5$). This could explain the high \al ratio 
observed in stars in
the thick disc and the halo of the Milky Way. 
This result do not depend strongly
upon the initial conditions, since it was found in all simulations.

Our main conclusion is that the morphological, kinematical and
chemical properties of the thick and thin disc can be reproduced in an scenario
where thick disc formed in a gas rich merger between two disc galaxies.
Furthermore, while the structural and kinematical 
properties of the merger remnants are strongly dependent on the initial
conditions --features such as the ratio of the disc scale-lengths,
the extent of the discs, the presence of rings
or of a counter-rotating component, vary from simulation to simulation--
the chemical abundances
show a remarkable consistency among the various simulations.
 The key result
of this study is that the ratio \al remains constant with increasing \fe for old stars,
up to $\rm[Fe/H]=-0.5$
while it decreases with increasing metallicity 
for young stars. This is true in each of the 8 simulations we considered. 
The observed  chemical signatures in our merger remnants require that mergers
happen before the onset of the majority of Type~Ia SNe, regardless of the orbits or mass ratios
of the progenitors. Furthermore, the mergers need to be gas rich to 
lead to a remnant with a rotationally supported disc. 
These characteristics are consistent with the mergers
we expect to find at high redshift and, therefore, our results support
this scenario for the formation of the different disc components 
in the Milky Way.

\section*{Acknowledgments}

The simulations were performed at the Laboratoire d'Astrophysique 
Num\'erique, Universit\'e Laval, the University of Central Lancashire's
High Performance Computing Facility, and
%on Cray XT4 
at the Center for Computational Astrophysics of the
National Astronomical Observatory of Japan.
The code used for generating the entries displayed in Table~\ref{angles}
was written by Keven Roy.
SR, \& HM acknowledge the support of the Canada Research Chair program 
and NSERC. PSB acknowledges the support of a Marie Curie Intra-European
Fellowship within the $6^{\rm th}$ European Community Framework
Programme. BKG and CBB acknowledge the support of the UK's
Science \& Technology Facilities Council (ST/F002432/1) and
the Commonwealth Cosmology Initiative.

\bsp

\label{lastpage}

\end{document}